\documentclass[12pt]{article}
\usepackage{amsfonts}
\usepackage{amsmath,amssymb}
\usepackage{stmaryrd}
\usepackage[dvips]{color}
\usepackage{fancybox}
\usepackage{bm}
\usepackage[dvips]{graphicx}
\usepackage{wrapfig}
\baselineskip=\normalbaselineskip

\begin{document} 

\numberwithin{equation}{section}
\renewcommand{\theequation}{\thesection.\arabic{equation}}
\def\natural{\mathbb{N}}
\newcommand{\Bra}[1]{\left\langle\, #1\,\right|}
\newcommand{\Ket}[1]{\left|\, #1\,\right\rangle}
\newcommand{\Bracket}[2]{\left\langle\, #1\,|\, #2\,\right\rangle}
\newcommand{\bracket}[1]{\left\langle\, #1\,\right\rangle}
\def\mat#1{\matt[#1]}
\def\matt[#1,#2,#3,#4]{\left(%
\begin{array}{cc} #1 & #2 \\ #3 & #4 \end{array} \right)}
\def\hq{\hat{q}}
\def\hp{\hat{p}}
\def\hx{\hat{x}}
\def\hk{\hat{k}}
\def\hw{\hat{w}}
\def\hl{\hat{l}}

\def\p{\partial}
\def\be{\begin{equation}}
\def\ee{\end{equation}}

\def\bea#1\ena{\begin{align}#1\end{align}}
\def\nn{\nonumber\\}
\def\cL{{\cal L}}
\def\TM{TM\oplus T^*M}
\newcommand{\CouB}[2]{\left\llbracket #1,#2 \right\rrbracket}
\newcommand{\pair}[2]{\left\langle\, #1, #2\,\right\rangle}

\null \hfill TU-949  \\[3em]
\begin{center}
{\LARGE \bf{D-brane on Poisson manifold and Generalized Geometry}}
\end{center}

\begin{center}
{T. Asakawa${}^\sharp$\footnote{
e-mail: asakawa@maebashi-it.ac.jp}, H. Muraki${}^{\flat}$\footnote{
e-mail: hmuraki@tuhep.phys.tohoku.ac.jp}, and S. Watamura${}^{\flat}$\footnote{
e-mail: watamura@tuhep.phys.tohoku.ac.jp}}\\[3em] 
${}^\sharp$
Department of Integrated Design Engineering,\\
Faculty of Engineering,\\
Maebashi Institute of Technology\\
Maebashi, 371-0816, Japan \\[1em]

${}^\flat$
Particle Theory and Cosmology Group \\
Department of Physics \\
Graduate School of Science \\
Tohoku University \\
Aoba-ku, Sendai 980-8578, Japan \\ [5ex]

\abstract{The properties of the D-brane fluctuations are investigated
using the two types of deformation of the Dirac structure, based on the
$B$-transformation and the $\beta$-transformation, respectively. 
The former gives the standard gauge theory with 2-form field strength.
The latter gives a non-standard gauge theory on the Poisson manifold
 with bivector field strength
 and the vector field as a gauge potential, where the gauge symmetry is a diffeomorphism generated by the Hamiltonian vector field. The map between the two
gauge theories is also constructed with the help of Moser's Lemma and the Magnus expansion. We also investigate the relation to the gauge theory on the noncommutative D-branes.
}
\end{center}

\vskip 2cm

\eject
\section{Introduction}

Among the various dualities in string theory, the T-duality is the most characteristic and also 
interesting one from the viewpoint of stringy geometry.
T-duality means that a closed string does not distinguish a pair of spaces, which are T-dual to each other. 
Since a string is the only object which observes the spacetime,
this property should be incorporated into the axioms of the stringy geometry.
It is also known that when we consider T-duality transformations under the existence of fluxes, 
many types of topologically nontrivial spaces appear. 
Among them, there are spaces which are not manifolds in a standard sense, 
generally called "non-geometric spaces". 
Since a closed string can travel in those "non-geometric spaces", 
they must be equally natural from the stringy geometry viewpoint.

Generalized geometry is a proposal by Hitchin to formulate a geometry
where these duality properties are realized manifestly \cite{Hitchin:2004ut,Gualtieri:2003dx}. 
Especially the generalized complex structure is useful  
to classify possible supersymmetric compactified spaces with fluxes such as Calabi-Yau manifolds.
Thus the generalized geometry is mostly applied to 
analyze the supergravity theories corresponding to the closed superstring theory \cite{Shelton:2006fd}.

However, it is also interesting how open strings and D-branes are characterized and 
behave in such a geometry \cite{Zabzine:2005qf, Koerber:2010bx}. 
For example, we know that a T-duality transformation changes the dimension of a
D-brane, and thus we expect that the effective theory 
based on the generalized geometry 
will treat D-branes of different dimensions in an unified way.

Recently, a geometrical characterization of D-branes in the framework of
 the generalized geometry has been proposed \cite{Asakawa:2012px}. 
There, a D-brane including fluctuations is identified with a 
leaf of a foliation generated 
by a Dirac structure of the generalized tangent bundle.
The scalar fields and vector fields on the D-brane are also unified
as a generalized connection. 
From this geometrical setting, the richer symmetry of the D-brane 
in the target space becomes transparent and it was shown that 
the Dirac-Born-Infeld (DBI) action, 
the effective action of the low energy theory of a D-brane,
realizes these symmetries nonlinearly. 

The nonlinear realization of the symmetry can be understood as a phenomenon following from the 
spontaneous symmetry breaking of a larger symmetry in the generalized geometry. 
The symmetry of a Dirac structure in the generalized geometry
is characterized by a foliation preserving generalized diffeomorphism, where
a leaf is mapped to another leaf keeping the foliation structure.
However, the existence of a D-brane chooses one of the leaves 
and thus breaks the symmetry to the leaf preserving generalized diffeomorphism,
which keeps the leaf itself.
Based on this picture, 
fields in the effective theory have been identified systematically 
as Nambu-Goldstone (NG) bosons associated with the broken symmetries,
and the nonlinear realizations of the broken symmetries are derived. 
These wider symmetries restrict the action of the
effective theory, and in the lowest order it is shown that the result is the DBI action.
Here the induced generalized metric appearing in the action is also understood as a 
generalized metric seen by the Dirac structure.

This result means that 
Dirac structures in the generalized geometry 
are the proper geometrical concepts to characterize D-branes.
In this paper, we develop this picture further.

Here we investigate the Dirac structure corresponding to 
a D-brane with a non-trivial gauge flux, or equivalently, 
a bound state of D-branes of various dimensions.
A typical property of such a situation is, as we explain below, 
that there are always two ways of describing the same Dirac structure.
For example, one can define a Dirac structure from $TM$ 
by using a so-called $B$-transformation generated by a symplectic 2-form $\omega$,
then a graph of the map corresponds to a Dirac structure which we call $L_\omega$.
On the other hand, the same graph can be described by a so-called $\beta$ transformation 
from the cotangent bundle $T^*M$, which is generated by a Poisson bivector $\theta$, 
and its graph becomes also a Dirac structure which we call $L_\theta$. 
$L_\omega$ and $L_\theta$ are dual descriptions of the same Dirac structure.
The first description $L_\omega$ is a direct generalization of the results 
given in a previous paper \cite{Asakawa:2012px}, 
and it fits to describe a D9-brane with a $2$-form gauge flux $\omega$.
Here we develop another formulation of a D-brane as a
Dirac structure based on the second description $L_\theta$.
As we will see, the proper language to formulate a gauge theory appearing naturally from this description is 
the differential calculus used in the Poisson cohomology,
where the role of $1$-forms and vector fields is exchanged.

Although the two Dirac structures $L_\omega$ and $L_\theta$ are equivalent, 
considering the fluctuations on the D-brane, we see that they are quite different.
In our previous paper we have shown that, including fluctuations, 
the D-brane can still be characterized by a Dirac structure.
In other words the fluctuation is identified with a deformation of the Dirac structure. 
Now the deformation can also be seen in two ways, either as a variation of the symplectic structure
$\omega'=\omega +\tilde{F}$ or a variation of the 
Poisson structure $\theta'=\theta +\hat{F}$.
Of course, the $2$-form $\tilde F$ and the bivector $\hat{F}$ are related with each other.
Then, the condition that the $L_{\theta'}$ obtained by a deformation becomes again a Dirac structure
is formulated by a Maurer-Cartan type equation for $\hat{F}$. 
As we will see, it is very natural to 
identify $\hat F$ with a kind of field strength and the Maurer-Cartan type equation 
with the Bianchi identity in a gauge theory.
We will find a gauge potential $\Phi$ corresponding to this "field strength" $\hat F$
 requiring that the Maurer-Cartan 
type equation is automatically satisfied, 
like the Bianchi identity is trivially satisfied if one writes the field strength 
$\tilde F$ by a gauge potential $a$ in a usual gauge theory. 

The above observation is also consistent 
with the analysis on the non-linear realization of broken symmetry 
for Dirac structures developed in \cite{Asakawa:2012px}.
There,
we found that the non-linearly realized symmetry 
 leads to the conclusion that the $2$-form $\tilde F$ can be written as 
a field strength of a gauge potential $a$,
and it describes an ordinary gauge theory on the $D$-brane with $U(1)$ gauge symmetry.
By using the same analysis, we arrive at the interesting conclusion that 
the bivector $\hat F$ can also be considered as a kind of field strength,
and the corresponding gauge potential is a vector field $\Phi$, 
and the gauge symmetry is a diffeomorphism generated by a set of Hamiltonian vector fields.

The new gauge theory is quite different in shape but, of course, 
it should be equivalent to the ordinary $U(1)$ gauge 
theory corresponding to $\tilde F$.
In fact, we find an explicit relation between the two gauge fields $a$ and $\Phi$, 
which is highly non-linear.
It is also shown that the two gauge theories are gauge equivalent, 
although the structures of gauge symmetries 
look quite different.
In the proof,  Moser's lemma \cite{Moser:1965}
which relates a deformation of a symplectic structure with a diffeomorphism, 
plays an important role.
We show that Moser's lemma is also explained in a natural way within the framework of generalized geometry.
We show that a diffeomorphism can be seen either as a $B$-gauge transformation or as a $\beta$-gauge transformation up to a generalized diffeomorphism that preserves the Dirac structure.
Another technique used to show this relation is the Magnus expansion \cite{Magnus:1954zz}, 
which relates a time ordered exponential to an ordinary exponential.

One motivation to study $L_\theta$ comes from a suggestion given in \cite{Jurco:2013upa},
in which $L_\theta$ is proposed to corresponds to a noncommutative description of a $D9$-brane 
in the $B$-field background.
Here we obtain in a quite simple way the two relations associated with a noncommutative D-brane:
the Seiberg-Witten (SW) relation between so-called open and 
closed string variables and the SW map between commutative and noncommutative gauge field strengths,
which is valid when all tensors are constant \cite{Seiberg:1999vs, Jurco:2001my}.
These relations are, however,  the result of two dual descriptions of the same Dirac structure
and valid in a wider context than merely the original noncommutative D-brane.

The paper is organized as follows.
In the next section,  we review the basic facts about generalized geometry.
In \S 3, we study how the two Dirac structures $L_\omega$ and $L_\theta$ characterize a D-brane.
We also reproduce two relations concerning noncommutative D-branes.
In \S 4, we focus on the differential geometry on the worldvolume.
We show that the Dirac structure is isomorphic to a Lie algebroid of a Poisson manifold.
In \S 5,
we argue that the two fluctuations $\tilde{F}$ and $\hat{F}$ are in fact field strengths,
from the viewpoint of both the Maurer-Cartan type relation and 
the non-linear realization of broken symmetries.
Then, we introduce the corresponding 
gauge potential and the gauge transformation.
After recalling Moser's lemma in \S 6, we obtain the relation between 
the two gauge potentials $a$ and $\Phi$ of $\tilde F$ and $\hat F$, respectively in \S 7. A short review on the Magnus expansion is also included there.
\S 8 is devoted to conclusions and discussions.


\section{Generalized geometry}

In this section, 
we present some basic facts about Lie algebroids \cite{CW} 
and the generalized geometry \cite{Hitchin:2004ut,Gualtieri:2003dx,Bouwknegt:2010zz},
with introducing the notations used in this paper.

\subsection{Lie algebroids}

First, we briefly introduce the notion of a Lie algebroid.
A Lie algebroid $(A, \rho, [\cdot ,\cdot]_A)$
consists of a vector bundle $A\rightarrow M$ over a base manifold $M$
together with a Lie bracket on its sections,
$[\, \cdot , \, \cdot \,]_A :\Gamma(A)\times \Gamma(A) \rightarrow \Gamma(A)$, and
an anchor map 
$\rho : A \rightarrow TM$ such that the induced map $\rho : \Gamma(A) \rightarrow \Gamma(TM)$ is a Lie-algebra homomorphism and the Leibniz identity
\begin{align}
	[X, fY]_A  = f [X,Y]_A + (\rho(X)\cdot f) Y,	\label{Leibniz}
\end{align}
is satisfied for any $X,Y \in \Gamma(A)$ and $f \in C^\infty (M)$.
Here $\rho(X)\cdot f $ denotes the action of the vector field $\rho(X)$ on $f$.
The tangent bundle $TM$ itself is a Lie algebroid with 
the anchor $\rho$ being the identity map.

\subsubsection{Differential algebra}

For a given Lie algebroid $(A,\rho,[\cdot,\cdot]_A)$, one can construct an algebra of associated 
$A$-differential forms $(\Gamma(\wedge^\bullet A^*), \wedge, d_A)$ in general.
Here the exterior differential $d_A$ is a map from a $k$-form $\omega \in \wedge^k A^*$
to a $(k+1)$-form $d_A \omega$, which is defined by
\bea
d_A \omega (X_1, \cdots, X_{k+1}) 
=&\sum_i (-)^{i+1}\rho (X_i)\cdot \omega (X_1, \cdots, \hat{X_i}, \cdots, X_{k+1})\nn
+&\sum_{i< j} (-)^{i+j}\omega ([X_i,X_j]_A, X_1, \cdots, \hat{X_i}, \cdots, 
\hat{X_j},\cdots, X_{k+1}).
\ena
As usual, $d_A$ is a graded derivation on a wedge product, 
and it satisfies $d^2_A=0$.

For the tangent bundle $A=TM$, it is the standard exterior differential 
algebra $(\Gamma(\wedge^\bullet T^*M), \wedge, d)$, where $d$ is the de Rham exterior differential.

\subsubsection{Gerstenhaber algebra}

Another algebra related to a Lie algebroid $A$ is the Gerstenhaber algebra 
$(\Gamma(\wedge^\bullet A), \wedge, [\cdot,\cdot]_A)$ 
of $A$-polyvectors.
Here the Gerstenhaber bracket is defined as a derived bracket, extending the Lie bracket.
For $V,W \in \wedge^\bullet A$,  it is 
\bea
i_{[V,W]_A}\omega =(-)^{(|V|-1)(|W|-1)} i_V d_A i_W \omega
-i_W d_A i_V \omega -(-)^{(|V|-1)}i_{V\wedge W}d_A \omega.
\label{Gerstenhaber bracket}
\ena
The bracket is graded commutative 
\bea
[V,W]_A =-(-1)^{(|V|-1)(|W|-1)}[W,V]_A,
\ena
and satisfies the graded Jacobi identity. 
Note that the bracket is defined in particular for $\wedge^0 A=C^\infty (M)$.
It is a extension of $A$-Lie derivative $[X,f]_A=\rho(X) \cdot f $ for $X \in A$ to any polyvector $V$.

In the case of $A=TM$, this bracket is called the Schouten(-Nijenhuis) bracket. 
For two polyvectors of the form $V=X_1\wedge \cdots \wedge X_k$ and $W=Y_1\wedge \cdots \wedge Y_l$
with $X_i, Y_j \in TM$,
the Schouten bracket in $\Gamma(\wedge^\bullet TM)$ is also written as the sum of Lie brackets of vector fields as
\bea
[V,W]_S=\sum_{i=1,j=1}^{k,l} (-1)^{i+j} [X_i,Y_j]\wedge X_1 \wedge \cdots \wedge \hat{X_i} \wedge \cdots 
X_k \wedge Y_1 \wedge \cdots \hat{Y_j}\wedge \cdots Y_l.
\label{def Schouten}
\ena
The bracket for $\wedge^0 A=C^\infty (M)$ is given for example $[X,f]_S=X\cdot f=X(df)$.

Under the anchor map $\rho:A\to TM$, 
its pullback $\rho^*: \Gamma(\wedge^\bullet T^*M) \to\Gamma(\wedge^\bullet A^*)$ 
is a morphism of differential algebras, that satisfies $\rho^* \circ d =d_A \circ \rho^*$.
On the other hand,  its natural extension 
$\wedge^\bullet \rho:  \Gamma(\wedge^\bullet A) \to\Gamma(\wedge^\bullet TM)$ 
is a morphism of Gerstenhaber algebras.

\subsection{Generalized tangent bundle}

Let $M$ be a $d$-dimensional smooth manifold, $TM$ be the tangent and
$T^\ast M$ be the cotangent bundle over $M$, respectively.
The sum of these bundles, ${\mathbb T} M=TM\oplus T^\ast M$ is 
called a generalized tangent bundle.
A section of ${\mathbb T} M$ is called 
a generalized vector field and will be represented by
a sum $X+ \xi$ of a vector field $X=X^\mu \p_\mu \in\Gamma(TM)$
and a 1-form $\xi=\xi_\mu dx^\mu \in \Gamma(T^\ast M)$.
It is equipped with the following operations:
\begin{enumerate}
\item  The anchor map $\pi$ is a bundle map ${\mathbb T}M\to TM$, the projection to the tangent bundle.
It induces a map $\Gamma({\mathbb T}M)\to \Gamma(TM)$ given by $\pi(X+\xi)=X$.
\item The canonical inner product is a fiberwise non-degenerate symmetric bilinear form 
$\left<\cdot,\cdot\right> : 
\Gamma({\mathbb T}M) \times\Gamma({\mathbb T}M) \rightarrow C^\infty (M)$, given by
\begin{equation}
\left< X+\xi , Y + \eta \right> = \frac{1}{2} (\imath_X \eta + \imath_Y \xi) 
=\frac{1}{2} 
\begin{pmatrix} X \\ \xi \end{pmatrix}^{\!\!T} 
\begin{pmatrix}0&1\\ 1&0 \end{pmatrix} 
\begin{pmatrix}Y \\ \eta \end{pmatrix},
\label{eq:canonical_inner_product:3kdo}
\end{equation}
where the last expression is written in terms of $2d$-component vectors.
\item The Dorfman bracket is a map $\Gamma({\mathbb T}M) \times\Gamma({\mathbb T}M)
\rightarrow\Gamma({\mathbb T}M)$ defined by
\begin{equation}
[X+\xi , Y+\eta] = [X, Y] + {\mathcal L}_X \eta - \imath_Y d\xi,
\label{eq:Dorfman_bracket1}
\end{equation}
where $[X,Y] \in \Gamma (TM)$
is the ordinary Lie bracket of vector fields, 
$\imath_X$ is the interior product , i.e. $\imath_X \eta = X^\mu \eta_\mu$ 
and ${\mathcal L}_X $ is the Lie derivative along a vector field $X$.
It is not skew-symmetric, but satisfies Jacobi identity.
\end{enumerate}
It is known that the bundle ${\mathbb T}M$ together with these structures 
is an example of Courant algebroids \cite{1990, Liu97manintriples}, 
which is a natural generalization of the Lie algebroid $TM$.
Note that the Courant bracket $[\cdot, \cdot]_C$, 
which is the skew-symmetrization of the Dorfman bracket (\ref{eq:Dorfman_bracket1}),
can also be used to define a Courant algebroid.
In the following, we do not distinguish a bundle ${\mathbb T} M$ and its sections 
$\Gamma({\mathbb T} M)$ if it is not confusing.
In this paper, we do not consider the $H$-twisted Courant bracket.

\subsubsection{Generalized Lie derivative}

In the ordinary differential geometry, 
a diffeomorphism $\varphi:M\to M$ induces an automorphism $\varphi_*: TM \to TM$ of the 
tangent bundle $TM$, so that the symmetry of the Lie algebroid $TM$ is the
diffeomorphism group ${\rm Diff}(M)$.
An infinitesimal diffeomorphism is generated by a vector field $\epsilon=\epsilon^\mu \p_\mu$, and 
its action on $TM$ is represented by the Lie derivative $\mathcal{L}_\epsilon $, as 
$\mathcal{L}_\epsilon =[\epsilon, X]$ for $X \in TM$. 

Similarly, the symmetry of the generalized tangent bundle is defined by the automorphism group
of the Courant algebroid ${\mathbb T}M$.
It is a semi-direct product ${\rm Diff}(M)\ltimes \Omega^2_{\rm closed}(M)$ 
of the ordinary diffeomorphism group and an abelian group of B-field 
transformations $\Omega^2_{\rm closed}(M)$, where the action of a closed 2-form $b$ on a section 
of ${\mathbb T}M$ is defined by $X+\xi \to X+\xi +i_X b$.
We call an element of this group as a generalized diffeomorphism. 

Infinitesimally, the Courant automorphism 
group defines the algebra of derivations, whose element is 
a generalized Lie derivative $\mathcal{L}_{(\epsilon ,b)}$ 
generated by a pair $(\epsilon ,b)$ of a vector field and a closed 2-form,
which acts on $X+\xi \in {\mathbb T} M$ as
\begin{align}
		\mathcal{L}_{(\epsilon ,b)}(X+\xi) := \mathcal{L}_\epsilon  (X+\xi)  + i_X b.
\end{align}
Especially when $b$ is exact, $b=-d\Lambda$, 
it reduces to the inner derivation with respect to the Dorfman
bracket in eq.(\ref{eq:Dorfman_bracket1}).
We denote the generalized Lie derivative $\mathcal{L}_{\epsilon+\Lambda}$ in such a case for brevity, then
and thus
\begin{align}
		 \mathcal{L}_{\epsilon+\Lambda}(X+\xi) 
=\mathcal{L}_\epsilon  (X+\xi)  - \iota_X d\Lambda 
=[\epsilon +\Lambda,X+\xi].
\label{inner derivation}
\end{align}

\subsection{Dirac structure}

A Dirac structure is a subbundle $L\subset \mathbb{T}M$, 
such that it is involutive under the Dorfman bracket $[X+\xi, Y+\eta] \in L$ for $X+\xi, Y+\eta \in L$,
it is isotropic under the canonical inner product 
$\big<X+\xi, Y+\eta \big>=0$ for $X+\xi, Y+\eta \in L$,
and it has the maximal rank. 
The Dorfman bracket restricted on $L$ is antisymmetric and thus 
a Dirac structure is a Lie algebroid by definition.

It is immediate to see that a generalized diffeomorphism (\ref{inner derivation}) 
generated by an element of $L$ is a symmetry of the Dirac structure $L$.
In fact the action $\cL_{\epsilon +\Lambda }$ for $\epsilon +\Lambda  \in L$ 
on a section $X+\xi \in L$ lies again in $L$, $\cL_{\epsilon +\Lambda }(X+\xi) \in L$,
because $L$ is involutive .
We call it as a $L$-diffeomorphism. 

Trivial examples of the Dirac structure are $TM$ and $T^\ast M$.
Less trivial examples are obtained by 
a $B$-transformation of $TM$ and a $\beta$-transformation of $T^\ast M$,
which we describe below.
They are the main objects studied in this paper.

\subsubsection{$B$-transformation of $TM$}

Given an arbitrary $2$-form $\omega \in \wedge^2 T^*M$,
a $B$-transformation of $TM$ defines a subbundle $L_\omega =e^{\omega} (TM)$,
\bea
	L_\omega= \{e^{\omega} (X)= X+\omega (X) \big| X\in TM	\}.
\ena
Here the 2-form $\omega$ is considered as a map $TM\rightarrow T^*M$, defined by
\be
\omega (X) :=\omega (X, \cdot) =\iota_X \omega 
=\omega_{\mu\nu} X^\mu dx^\nu,
\ee
where the last term is a local expression written in the components of
$\omega=\frac{1}{2}\omega_{\mu\nu}dx^\mu\wedge dx^\nu$.
The subbundle $L_\omega$ is a Dirac structure if and only 
if $\omega$ is a closed 2-form, $d\omega=0$. 
This is because the $B$-transformation generated by a closed $2$-form is a symmetry
of the generalized tangent bundle.

\subsubsection{$\beta$-transformation of $T^\ast M$}

Given an arbitrary bivector $\theta \in \wedge^2 TM$,
a $\beta$-transformation of $T^*M$ defines a subbundle $L_\theta =e^{\theta} (TM)$,
\bea
	L_\theta= \{e^{\theta} (\xi)= \xi+\theta (\xi) \big| \xi\in T^*M	\}.
\ena
Here, the bivector $\theta$ is considered as a map $T^*M\rightarrow TM$, defined by
\be
\theta (\xi) :=\theta (\xi,\cdot) =\iota_\xi \theta =\theta^{\mu\nu} \xi_\mu \p_\nu,
\ee
where the last term is a local expression written in components of
$\theta=\frac{1}{2}\theta^{\mu\nu}\p_\mu\wedge \p_\nu$.
The subbundle $L_\theta$ is a Dirac structure if and only if 
$\theta$ is a Poisson bivector, i.e., $[\theta, \theta]_S =0$
where $[\,\cdot,\,\cdot]_S$ is the Schouten bracket (\ref{def Schouten}) 
of polyvector fields. 
The condition to be a Poisson bivector can be written as
$\theta^{\mu\tau}\partial_\tau \theta^{\nu\rho}+\theta^{\nu\tau}\partial_\tau \theta^{\rho\mu}+\theta^{\rho\tau}\partial_\tau \theta^{\mu\nu}=0$ in components, 
which is the same condition for 
the Jacobi identity of the Poisson bracket 
$\{f,g\}=\theta(df,dg)$ for $f,g \in C^\infty (M)$. 

Note that a $\beta$-transformation is not a symmetry of the generalized tangent bundle ${\mathbb T}M$.
As a result, $L_\omega$ and $L_\theta$ are different in many respects,
that we study in this paper. 
See more information on $L_\theta$ as a Lie algebroid in \S 4 and 
a proof of the condition in Appendix.

\section{D-branes and Dirac structures}

As we have addressed in the introduction, a D-brane can be characterized 
by a Dirac structures \cite{Asakawa:2012px}.
The worldvolume of the D-brane is identified with a leaf of foliation, 
an integral manifold determined by the image of the anchor map $\pi:L \to TM$.
The set of all possible leaves just means that of all possible positions of a D-brane (moduli space)
obtained by transverse displacements.
A spacetime with a D-brane is thus a foliated manifold that admit such a foliation. 
In particular, a Dirac structure $L=TM$ corresponds to a spacetime filling D-brane,
i.e. $D9$-brane in superstring theory.

In this section, we assume that the Dirac structures $L_\omega$ and $L_\theta$ 
introduced in the previous section are also giving characterizations of certain kinds of D-branes. 
As a consequence, there are indications that $L_\theta$ corresponds to a noncommutative D-brane.

\subsection{D-brane as a leaf of foliation}

Let us start our discussion with a Dirac structure $L_\omega =e^\omega (TM)$ with a closed $2$-form $\omega$.
Since the anchor map $\pi: L_\omega \to TM$ is surjective, we understand that the world volume of corresponding D-brane 
is the base manifold $M$ and thus the D-brane is space-time filling.
Since the Dirac structure $L_\omega$ is a $B$-transformation of $TM$, $\omega$ is identified with 
a non-trivial $U(1)$-gauge flux on a $D9$-brane.
The existence of a $U(1)$-gauge flux 
generates lower D-brane charges (brane within brane),
it is equivalent to a bound state of $D9$-brane and lower dimensional D-branes.

The case where the Dirac structure is 
specified by $L_\theta =e^\theta (T^*M)$ is more subtle.
According to the identification above, 
the world volume is a symplectic leaf of a Poisson manifold $(M,\theta)$, 
which is a symplectic manifold but whose dimension depends on the rank of the bivector 
$\theta$ at that point.
Then the spacetime is a collection of symplectic leaves with various dimensions.
To avoid this complexity, we focus on the nondegenerate Poisson structure (full rank everywhere)
in this paper.
Correspondingly, $L_\theta$ is identified with a kind of $D9$-brane, 
whose worldvolume is a nondegenerate Poisson manifold $(M,\theta)$.

\subsection{Two ways to describe the same Dirac structure}

The above assumption is equivalent to the requirement that the spacetime $(M,\omega)$ is a symplectic manifold
where $\omega$ is a symplectic structure (i.e., non-degenerate).
In the generalized geometry, 
it means that there are two possibilities $L_\theta$ and $L_\omega$ which define the same Dirac structure, as we will see.

\begin{figure}[htbp]
\begin{center}
\includegraphics[width=0.4\textwidth]{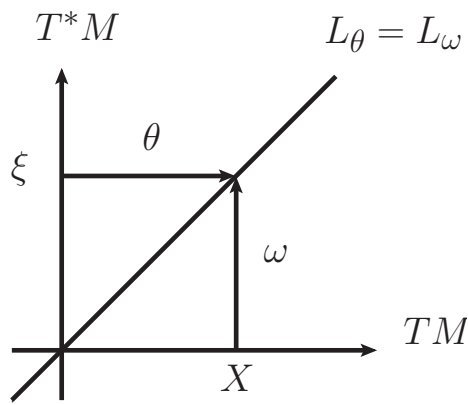}
\caption{The relation between $L_\omega$ and $L_\theta$}
\label{fig:trick}
\end{center}
\end{figure}

Any element of $L_\omega$ can be represented by using a vector $X\in TM$ as 
$X+\omega(X)$, 
and any element of $L_\theta$ can be represented by using 1-form 
$\xi\in T^*M$ as $\xi+\theta(\xi)$. 
If two Dirac structures are the same subbundle $L_\theta=L_\omega$, 
there must be one to one correspondence between these two representations.
Thus we have (see Fig.\ref{fig:trick})
\be
\xi+\theta(\xi)=X+\omega(X).
\ee
Comparing the vectors and forms 
in both sides, 
we get
\be
\xi=\omega(X)~,~X=\theta(\xi).
\ee
The first relation defines a 1-form $\xi$ for a vector $u$,
and then the second equation gives a consistency
\be
X=\theta(\omega(X)),
\ee
which should be satisfied for an arbitrary $X$.
In components this gives 
a relation of matrices $\theta^{\mu\nu}=(\omega_{\mu\nu})^{-1}$.
For brevity, we write it as
\be
\theta=\omega^{-1}.
\ee

This is a simple demonstration of a trick 
which we are going to use frequently in the following sections.
In this example, it gives a rather trivial statement 
that a symplectic structure defines a Poisson bivector as its inverse.
However this "trick" is useful to identify two subbundles defined by the 
different ways in more complicated cases. 

We give two more such examples below, and interestingly, both of them suggest the connection between 
a Dirac structure $L_\theta$ and a noncommutative description of a D-brane.

\subsection{Metric seen by $L_\theta$}

Here, we want to discuss a metric structure on the D-brane in the present formulation. 
For this purpose we need to introduce a generalized Riemannian structure \cite{Gualtieri:2003dx}.
The generalized Riemannian structure $C_+ \subset {\mathbb T} M$ is a subbundle, 
which is defined as a set of positive-definite 
generalized vectors $V=X+\xi$, s.t. $\bracket{V,V}>0$,
and can be represented by a graph of a map of the generalized metric
 $E=g+b : TM\to T^*M$, i.e.,
\begin{align}
C_+ =\{X+E(X) ~|~ X\in TM\}.
\end{align}
where $g$ and $B$ are the Riemannian metric and the $B$-field, respectively.
Note that the generalized metric tensor $E=g+b$ comes from the closed string background, 
and it is independent of the Dirac structure 
related with a D-brane which originates from the open strings.

As argued in \cite{Asakawa:2012px}, 
we can represent the same $C_+$ by a graph of a different map 
$t: L\to L^*$, from an arbitrary Dirac structure $L$
 to its dual $L^*$,
 instead of the map from $TM$ to $T^*M$.
In the case of Dirac structures treated in \cite{Asakawa:2012px}, 
the map $t$ is identified with the induced generalized metric on the Dp-brane,
corresponding to $L$.
We call such $t$ as a metric seen from $L$.

We consider here a map $t: L_\theta \to L^*_\theta=TM$, 
which maps any element $\xi+\theta(\xi)$ in $L_\theta$ to a vector $t(\xi)$
and define a graph of the Riemannian structure $C_+$.
Then any section of $C_+$ is given by
\begin{align}
C_+ =\{\xi+\theta(\xi)+t(\xi) ~|~ \xi \in T^*M\}.
\end{align}
\begin{figure}[htbp]
\begin{center}
\includegraphics[width=0.4\textwidth]{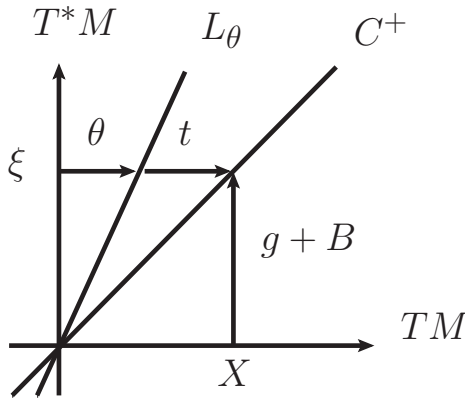}
\caption{The generalized Riemannian structure $C_+$ seen from $L_\theta$ and $TM$}
\label{fig:metric}
\end{center}
\end{figure}
Since the above two graphs 
define the same Riemannian structure $C_+$,
there must be one to one correspondence (see Fig.\ref{fig:metric})
\begin{equation}
X+E(X)=\xi+\theta(\xi)+t(\xi),
\end{equation}
or equivalently,
\begin{equation}
X=\theta(\xi)+t(\xi), ~~ E(X)=\xi.
\end{equation}
As a result, we have a relation 
\begin{align}
E^{-1}=\theta +t.
\label{NC metric}
\end{align}
If we write $t=(G+\Phi)^{-1}$, 
where $G$ is a symmetric tensor and $\Phi$ is an anti-symmetric tensor, 
then (\ref{NC metric}) is nothing but the relation given in \cite{Seiberg:1999vs}:
\begin{align}
\frac{1}{g+B}=\frac{1}{G+\Phi}+\theta.
\label{Seiberg}
\end{align}
This is usually called the Seiberg-Witten relation, 
or the open-closed relation,
in the case that all tensors are constant.
In the right hand side, $\theta$ is a noncommutative parameter of the D-brane,
$G$ is a metric seen by the noncommutative D-brane, 
and $\Phi$ is a background $2$-form.  
This suggests that the Dirac structure $L_\theta$ corresponds 
to a noncommutative D-brane \cite{Jurco:2013upa}.
Note that in this formulation, 
all the tensors are not restricted to be constant.
Note also that it is natural to regard the Poisson bivector 
$\theta$ as a free 
parameter in the relation, 
and it is simply associated with a choice of a Dirac structure $L_\theta$, i.e., a choice of 
the description of the D-brane.
On the other hand, the remaining tensors $g+B$ and $G+\Phi$ are 
background fields coming from closed strings.

The generalized Riemannian structure $C_+$ can also be seen from $L_\omega$.
In this case, the relation corresponding to (\ref{Seiberg}) is rather trivial 
$g+B= t+\omega$, where $t=g+B-\omega$ is the generalized metric seen by $L_\omega$.
Here, the 2-form $\omega$ is a free parameter.

\subsection{Fluctuations and SW map}

If we regard both structures $L_\theta =L_\omega$ as a kind of $D9$-branes, 
there are two natural candidate of D-brane fluctuations 
corresponding to the choice $L_\omega$ and $L_\theta$, respectively.
The validity of these fluctuations from a gauge theory point of view will 
be studied concretely in the next sections.
Here we focus on the relation of the two fluctuations 
using the same method in the previous subsection. 

Starting from $L_\omega$, its $B$-transformation by an arbitrary 
$2$-form $\tilde{F}=\frac{1}{2}F_{\mu\nu}dx^\mu\wedge dx^\nu \in \wedge^2 T^*M$
defines a new subbundle $L_{\omega+\tilde{F}}$, in which the element has a form
\bea
e^{\tilde{F}} (X+\omega (X))=X + (\omega +\tilde{F}) (X).
\ena 
This is a natural way to add a fluctuation for the D-brane 
characterized by $L_\omega$.

Another way is obtained by a $\beta$-transformation of $L_\theta$ by a
bivector $\hat{F}=\frac{1}{2}\hat{F}^{\mu\nu}\p_\mu\wedge \p_\nu \in \wedge^2 TM$.
It defines a new subbundle $L_{\theta +\hat{F}}$, in which an element has a form
\bea
e^{\hat{F}} (\xi+\theta (\xi ))=\xi + (\theta +\hat{F} )(\xi ).
\ena

\begin{figure}[htbp]
\begin{center}
\includegraphics[width=0.4\textwidth]{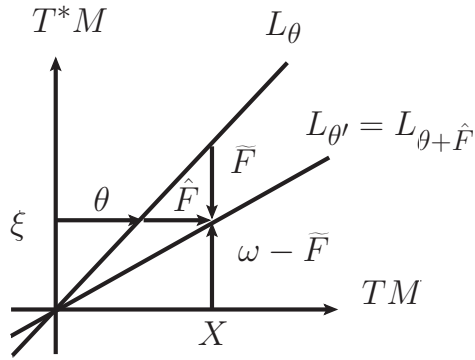}
\caption{The relation between $\tilde{F}$ and $\hat{F}$}
\label{fig:SWmap}
\end{center}
\end{figure}

We want to consider that the above two deformed subbundles are the
same D-brane system including the fluctuation and thus the two  
subbundles should be equivalent $L=L_{\omega+\tilde{F}}=L_{\theta +\hat{F}}$,
and thus their sections are relating each other as (see Fig.\ref{fig:SWmap})
\bea
&\xi + (\theta +\hat{F} )(\xi )=X + (\omega +\tilde{F}) (X),
\ena
or equivalently, 
\bea
(\theta +\hat{F} )(\xi )=X, ~~ \xi =(\omega +\tilde{F}) (X).
\ena
This condition leads to a matrix relation $(\theta +\hat{F} )(\omega +\tilde{F}) =1$, 
corresponding to the relation $\theta \omega=1$ for the case without fluctuation.
Therefore, we have
\bea
\theta +\hat{F}=\frac{1}{\omega +\tilde{F}}=\frac{1}{1+\theta \tilde{F}}\theta
\ena
and thus
\bea
\hat{F}=-\frac{1}{1+\theta\tilde{F}}\theta \tilde{F}\theta.
\ena
This relation is the same as the Seiberg-Witten map between 
commutative and noncommutative $U(1)$ field strength, 
if the $2$-form $\omega \hat{F} \omega$ is identified with a noncommutative field strength, and
in the case that all of $\theta$, $\tilde{F}$, and $\hat{F}$ are constant tensors \cite{Seiberg:1999vs}.
It suggest again a connection to noncommutative gauge theories \cite{Jurco:2013upa, Jurco:2001my}.
However,  our relation is valid for more general tensors.
Moreover, it is not yet clear whether 
$\tilde{F}$ and $\hat{F}$ are considered to be field strengths of some gauge theories.
In the remainder of this paper, we elaborate on the latter question.
First, in the next section, we study $L_\omega$ and $L_\theta$ themselves, 
focusing on the differential geometry on the worldvolume of the D-brane.
Then, in the subsequent section, we argue that $\tilde{F}$ and $\hat{F}$ are indeed gauge field strengths.



\section{Differential geometry on worldvolume}

A Dirac structure is automatically a Lie algebroid, 
and there is a proper differential calculus associated with a Lie algebroid.
In this section, we focus on such calculi associated with $L_\omega$ and $L_\theta$,
as a preparation for formulating gauge theories.
In particular, we see that $L_\theta$ is the same as a Lie algebroid of a Poisson manifold, 
and thus the calculus is that used for the Poisson cohomology.

\subsection{Lie algebroid of Poisson manifold}

First we introduce a Lie algebroid associated with a Poisson manifold, 
following \cite{CW}.
Let $(M,\theta)$ be a Poisson manifold, with a Poisson bivector $\theta$.
Then a Lie algebroid $A=(T^*M)_\theta$, or more precisely,   
$(T^*M, \theta, [\cdot,\cdot]_K)$
is defined as follows.
The anchor map $\rho: T^*M \to TM$
of the algebroid $(T^*M)_\theta$ is given by the Poisson bivector $\theta$.
Here the Poisson bivector $\theta =\frac{1}{2}\theta^{\mu\nu}\partial_\mu\wedge \partial_\nu \in \wedge^2 T^*M$ defines a map $\theta : T^*M \to  TM$  
by
$\theta(\xi)=i_\xi\theta=\theta^{\mu\nu}\xi_\mu \partial_\nu$ for $\xi\in T^*M$.
The Lie bracket of $A$ is given by so-called Koszul bracket, which 
is defined for $\xi,\eta\in T^*M$ by
\bea
[\xi,\eta ]_K &:=\cL_{\theta (\xi )}\eta -\cL_{\theta (\eta)}\xi-d (\theta (\xi,\eta ))\nn
&=i_{\theta (\xi )}d\eta -i_{\theta (\eta )}d\xi+d (\theta (\xi,\eta ))
\label{Koszul}
\ena
and in components
\bea
[\xi,\eta ]_K  
&=\theta^{\mu\nu} \xi_\mu \partial_\nu \eta_\rho dx^\rho
+\theta^{\mu\nu} (\partial_\mu \xi_\rho) \eta_\nu   dx^\rho
+(\partial_\rho \theta^{\mu\nu}) \xi_\mu \eta_\nu dx^\rho.
\label{Poisson Lie bracket in components}
\ena
The Koszul bracket reduces for exact 1-forms to the Poisson bracket as
\bea
[df,dg]_K=d(\theta (df, dg))=d\{ f,g\}.
\ena
We refer to this as the Lie algebroid of the Poisson manifold \cite{CW}.

As explained in \S2, for a given Lie algebroid $(A,\rho,[\cdot,\cdot])$, one can construct 
the exterior differential algebra $(\Gamma(\wedge^\bullet A^*), \wedge, d_A)$ and 
the Gerstenhaber algebra 
$(\Gamma(\wedge^\bullet A), \wedge, [\cdot,\cdot]_A)$.

In our case of $A=(T^*M)_\theta$, the 
differential algebra $(\Gamma(\wedge^\bullet TM), \wedge, d_\theta)$
is that of polyvector fields.
For a vector $X \in TM$, we have 
\bea
d_\theta X(\xi,\eta) 
&=\theta (\xi) (X(\eta)) -\theta (\eta)(X(\xi)) -X([\xi,\eta]_K)\nn
&=(\cL_{\theta (\xi)} X)(\eta) -(\cL_{\theta (\eta)} X)(\xi) 
+X(d (\theta (\xi,\eta ))\nn
&=[\theta, X]_S (\xi,\eta),
\ena
where in the last line we used the Schouten bracket for polyvector fields.
In general, 
\bea
d_\theta =[\theta, \cdot ]_S.
\ena
It is nilpotent $d_\theta^2=\frac{1}{2}[[\theta,\theta]_S, \cdot]_S=0$ 
because $\theta$ is a Poisson structure 
$[\theta,\theta]_S=0$.
The cohomology group defined with respect to $d_\theta$ is called the Poisson cohomology. 

On the other hand, the Gerstenhaber bracket in the algebra 
$\Gamma(\wedge^\bullet T^*M)$ is the extension of the Koszul bracket 
of $1$-forms defined as in (\ref{def Schouten}).

\subsection{Lie algebroid of Poisson manifold as Dirac structure}

Here we see that the Lie algebroid $A=(T^*M)_\theta$ is isomorphic 
to the Dirac structure $L_\theta=e^\theta (T^*M)$, as shown in \cite{Liu97manintriples}.

Since any Dirac structure is a Lie algebroid, the Courant bracket of the Dirac structure 
$L_\theta$ is automatically a Lie bracket (This is also true for the Dorfman bracket.).
The Courant bracket for two elements in $L_\theta$ is
\bea
[\xi+\theta(\xi), \eta+\theta(\eta)]_C
&= [\theta(\xi),\theta(\eta)]+\cL_{\theta(\xi)}\eta -\cL_{\theta(\eta)}\xi
-\frac{d}{2}(i_{\theta(\xi)}\eta-i_{\theta(\eta)}\xi)\nn
&=[\theta(\xi),\theta(\eta)]+\cL_{\theta(\xi)}\eta -\cL_{\theta(\eta)}\xi
-d(\theta(\xi,\eta)).
\ena
The $T^*M$-part is nothing but the Koszul bracket $[\xi,\eta]_K$  of
the Lie algebroid $A=(T^*M)_\theta$, defined in (\ref{Koszul}).
Similarly, the $TM$-part can be rewritten as (see Appendix for the proof.)
\bea
[\theta(\xi),\theta(\eta)]=\theta([\xi,\eta]_K),
\label{key relation for Poisson Dirac}
\ena
and thus the Courant bracket of the two elements in $L_\theta$ is 
again of the form
\bea
[\xi+\theta(\xi), \eta+\theta(\eta)]_C 
&=\theta([\xi,\eta]_K) +[\xi,\eta]_K.
\label{Poisson Dirac iso}
\ena
Thus, two Lie algebroids  $(T^*M)_\theta$ and $L_\theta$ can be identified with each other.
More precisely, a bundle map $\iota: A \to L_\theta$ such that 
$\iota(\xi)=\xi+\theta(\xi)$ is an isomorphism of Lie algebroids.

\subsection{Worldvolume of the D-brane }

As was explained already, 
the D-brane worldvolume associated with the Dirac structure $L_\theta$ is a Poisson manifold $(M,\theta)$.
From the isomorphism above, we may say that the worldvolume is also associated with the Lie algebroid $(T^*M)_\theta$.
The difference of the representations can be interpreted as follows:
the Lie algebroid $(T^*M)_\theta$ corresponds to a worldvolume before embedded into 
a spacetime, and the Dirac structure $L_\theta$ is a image of the embedding map.
In any case, it is clear that a natural differential geometry on the worldvolume,
to formulate a gauge theory on it, is the differential algebra $(\Gamma(\wedge^\bullet TM), \wedge, d_\theta)$
as well as the Gerstenhaber algebra $(\Gamma(\wedge^\bullet T^*M), \wedge, [\cdot,\cdot]_S)$.
In particular, in the differential geometry based on $(T^*M)_\theta$, 
the roles of $1$-forms and vector fields are exchanged, as compared to the ordinary differential geometry 
based on $TM$.
As a consequence, a gauge theory on a Poisson manifold $(M,\theta)$ 
is naturally described in terms of polyvectors.
This is indeed the case, which will become clear in the next section.

Similarly, the D-brane worldvolume associated with the Dirac structure $L_\omega$ 
is a symplectic manifold $(M,\omega)$.
In this case, the similar isomorphism between the Lie algebroid $TM$ and $L_\omega$ is given by the map $X \to X+\omega(X)$.
We can use the ordinary differential calculus based on the de Rham differential forms.
It means that we can apply the same calculus for the Dirac structure $L_\omega$ corresponding to a bound state of D-branes and for the Dirac structure $TM$ corresponding to a single $D9$-brane.

As we stated, we can take both $L_\omega$ and $L_\theta$ as a Dirac structure to describe the same D-brane.  However, the differential calculus is 
quite different in the two representation.
For the case with the symplectic form $\omega$, we can put $\omega$ on the worldvolume $M$ as a field and the differential calculus is not modified. 
On the other hand, for the case with the bivector $\theta$, 
the existence of the bivector is essential for the 
differential geometry on the worldvolume.
Thus, when $\theta=\omega^{-1}$, we have the two quite different descriptions of the same D-brane bound state.

We recall that in the paper \cite{Jurco:2001my}, 
polyvectors were used to describe a gauge theory on a Poisson manifold,
as a first step to formulate a noncommutative gauge theory.
However, as we will see, the gauge field in this paper is different from theirs.


\section{Fluctuations from broken symmetries}

In the analysis performed in \cite{Asakawa:2012px}, the symmetry preserved by 
the Dirac structure and the symmetry spontaneously broken by putting a D-brane
are important to identify the correct fluctuations.
We take here the same strategy for the Dirac structure $L_\theta$ as well as $L_\omega$.
In particular for $L_\theta$, an exotic type of a gauge potential will be obtained, corresponding to 
the differential geometry on a Poisson manifold.

\subsection{The case with $L=TM$}

First we briefly formulate the case of $L=TM$ \cite{Asakawa:2012px}.
Among the total symmetry ${\rm Diff} (M) \ltimes \Omega^2_{\rm closed} (M)$ of the generalized 
tangent bundle (Courant algebroid) ${\mathbb T} M$, 
the symmetry preserved by the Dirac structure $TM$ consists of the diffeomorphism 
generated by $\epsilon =\epsilon^\mu \partial_\mu$ 
and the $B$-field gauge transformation generated by
closed $1$-form $\Lambda$ ($d\Lambda=0$).
In the case of $M={\mathbb R}^D$, the latter means that $\Lambda=d\lambda $ for a function $\lambda$, 
so that the unbroken symmetry is ${\rm Diff} (M) \ltimes U(1) \ni \epsilon +d\lambda$.
On the other hand, the broken symmetry consists of $B$-field gauge transformations generated by
non-closed $1$-form $A=A_\mu dx^\mu$, which is in general a Nambu-Goldstone (NG) boson.
It acts on $X \in TM$ by the generalized Lie derivative ${\cal L}_{A}$ as 
\bea
e^{-{\cal L}_{A}}X =e^{dA}X=X+F(X), \quad F=dA.
\ena
Thus $A$ produces a new Dirac structure $L=L_{F}$, and 
is regarded as a gauge-field fluctuation on the $D9$-brane.
The above $U(1)$ symmetry is nothing but the gauge transformation $A\to A+d\lambda$.

\subsection{Ordinary gauge field}

Now we consider the case of $L_\omega$ of the form
\bea
L_\omega =\{ X+\omega (X) | X\in TM\}.
\ena
Note that its dual bundle is $L_\omega^*=T^*M$ and thus, we may write 
$TM\oplus T^*M=L_\omega \oplus L_\omega^*$.

Correspondingly, let us take a parametrization of total symmetry 
${\rm Diff} (M) \ltimes \Omega^2_{\rm exact}(M)$ as
$\epsilon + \omega(\epsilon )+\Lambda$, where $\epsilon \in TM$ and $\Lambda \in T^*M$.
It acts on a section $X+\omega (X) \in L_\omega $ as a generalized Lie derivative as 
\bea
{\cal L}_{\epsilon + \omega(\epsilon )+\Lambda}(X+\omega (X))
&={\cal L}_{\epsilon}X +\omega ({\cal L}_{\epsilon }X) -i_X d\Lambda .
\ena
Clearly, a transformation
generated by a generalized vector $\epsilon + \omega(\epsilon )$ 
is an unbroken symmetry for any $\epsilon \in TM$,
since it preserves $L_\omega$.
We call this symmetry as $L_\omega$-diffeomorphism ($L_\omega$-diff. for short).
In addition, a $B$-field gauge transformation generated by
a closed $1$-form $d\Lambda=0$ is also a symmetry.
This includes $U(1)$ with $\Lambda=d\lambda$ for any $\lambda \in C^\infty (M)$.
Therefore, the unbroken symmetry of $L_\omega$ is ${\rm Diff} (M) \ltimes U(1)$ for $M={\mathbb R}^D$, which is the same as that of $TM$.

On the other hand, the broken symmetry is a $B$-field gauge transformation generated by
a non-closed $1$-form $a=a_\mu dx^\mu \in T^*M$, which is a NG-boson.
It is again a $U(1)$ gauge potential with a gauge transformation $a\to a+d\lambda$.
Note that $a \in L_\omega^*$ is regarded as a generalized connection on $L_\omega$.
This broken symmetry produces a new Dirac structure $L=L_{\omega +\tilde{F}}$ as
\bea
e^{-{\cal L}_{a}} (X+\omega (X))
&=X+(\omega +\tilde{F})(X),
\ena
where $\tilde{F}=da \in \wedge^2 T^*M$ is the corresponding $U(1)$ field strength.
This is nothing but the fluctuation $\tilde{F}$ discussed in \S 3.4.
The condition for the Dirac structure $d\tilde{F}=0$ is nothing but the Bianchi identity from the
 gauge theory point of view, and is trivially satisfied by $\tilde{F}=da$ 
(of course, we can also consider a non-trivial $U(1)$ gauge flux $\tilde{F}$
by replacing $a$ with a set of locally defined $1$-forms, as usual.).
We can say that the new symplectic structure 
$\omega'=\omega+\tilde{F}$ is in the same class as $\omega$ 
in the second de Rham cohomology group $[\omega']=[\omega] \in H_{\rm de Rham}^2(M)$.

\subsection{New type of gauge field}

Now let us consider the Dirac structure $L_\theta$. The definition is 
\bea
L_\theta =\{ \xi +\theta (\xi) | \xi \in T^*M\}.
\ena
The dual bundle is $L_\theta^*=TM$ and thus, we can write 
$TM\oplus T^*M=L_\theta \oplus L_\theta^*$.

Correspondingly, we adopt a parametrization of the total symmetry ${\rm Diff} (M) \ltimes \Omega^2_{\rm exact}(M)$ as
$\epsilon + a + \theta (a)$, where $\epsilon \in TM$ and $a \in T^*M$.
The action of an infinitesimal transformation on a section $\xi +\theta (\xi) \in L_\theta$ can be represented by a generalized Lie derivative as
\bea
{\cal L}_{\epsilon + a + \theta (a)}(\xi +\theta (\xi) )
&=({\cal L}_{\epsilon + \theta (a)}\xi -i_{\theta (\xi )}da) 
+\theta ({\cal L}_{\epsilon + \theta (a)}\xi -i_{\theta (\xi )}da)
+({\cal L}_{\epsilon} \theta) (\xi).
\ena
From this expression, it is evident that the $L_\theta$-diff., 
the transformation
generated by $a + \theta (a)$ is an unbroken symmetry for any $a \in T^*M$.
In addition, as we see from the last term in the r.h.s., a $\theta$-preserving diffeomorphism generated by
a vector field $\epsilon \in TM$ such that ${\cal L}_\epsilon \theta=0$ is also a symmetry.
This class of transformations includes the diffeomorphism generated by the Hamiltonian vector fields 
$\epsilon =X_f =\theta (df)$ for any $f \in C^\infty (M)$.

On the other hand, the rest of the diffeomorphisms is broken.
Denote a generator of those diffeomorphisms 
as $\Phi =\Phi^\mu \partial_\mu \in TM$.
It is a NG-boson according to the general argument of the broken symmetries.
The vector field $\Phi$ can be also seen as a kind of the 
gauge field, since it is a generalized connection on $L_\theta$, 
i.e., $\Phi \in TM =L_\theta^*$. 
The infinitesimal action of the vector field $\Phi$ on $L_\theta$ is
\bea
{\cal L}_{\Phi}(\xi +\theta (\xi) )
&={\cal L}_{\Phi}\xi
+\theta ({\cal L}_{\Phi}\xi )
+({\cal L}_{\Phi} \theta )(\xi),
\ena
and the finite action is 
\bea
e^{-{\cal L}_{\Phi}}(\xi +\theta (\xi) )
&=e^{-{\cal L}_{\Phi}}\xi
+(e^{-{\cal L}_{\Phi}}\theta )(e^{-{\cal L}_{\Phi}}\xi ).\label{Phi transformation}
\ena
By defining $\xi'=e^{{\cal L}_{\Phi}}\xi $, and $\theta'=e^{-{\cal L}_{\Phi}}\theta$,
the r.h.s. of (\ref{Phi transformation}) is rewritten as $\xi' +\theta' (\xi' ) $, which is a section of a new subbundle $L_{\theta'}$.
Therefore, the diffeomorphism by $\Phi $ causes a map of subbundles $L_\theta \to L_{\theta'}$,
which is effectively seen as a $\beta$-transformation
generated by a bivector $\hat{F}\in \wedge^2 TM$ defined by
\bea
\theta' =\theta + \hat{F}.
\ena
(We will return to more systematic analysis on the relation between 
diffeomorphisms and $\beta$-transformations in \S 6.)
In other words, $\hat F$ is represented by the field $\Phi$ as
\bea
\hat{F} =e^{-{\cal L}_{\Phi}}\theta -\theta.
\label{F via Phi}
\ena
We now argue that $\hat{F}$ is a generalized field strength,
corresponding to a generalized connection $\Phi$.

\subsubsection{$\hat{F}$ as field strength}
 
Let us consider the $\beta$-transformation $e^{\hat{F}}$ of the Dirac structure $L_\theta$,
which defines a new subbundle $L_{\theta'}$ with $\theta'=\theta+\hat{F}$.
The deformed subbundle $L_{\theta'}$ stays in a Dirac structure if and only if $\theta'=\theta+\hat{F}$ is a Poisson structure. This requires 
\be
0=[\theta',\theta']_S=[\theta +\hat{F},\theta+\hat{F}]_S=
[\theta,\theta]_S+2[\theta, \hat{F}]_S+[\hat{F},\hat{F}]_S
\ee
Thus, $\hat{F}$ should satisfy the Maurer-Cartan-type equation \cite{Liu97manintriples}:
\be
d_\theta \hat{F}+\frac{1}{2}[\hat{F},\hat{F}]_S =0.\label{MCtypeeq}
\ee
Here we used the notion of $d_\theta$ and the Schouten bracket in the differential geometry discussed in the previous section.

When we consider $\hat{F}$ as a field strength, this condition play the role of the Bianchi identity.
It means that (\ref{MCtypeeq}) should be trivially satisfied, when $\hat{F}$ is written in terms of the corresponding gauge potential.
In fact, for an arbitrary vector field $\Phi \in TM$, the $\hat F$ in (\ref{F via Phi}) 
satisfies (\ref{MCtypeeq}) automatically.
It is easily shown by noting
\bea
d_\theta \hat{F}&=d_\theta (e^{-{\cal L}_\Phi}\theta ),\nn
\frac{1}{2}[\hat{F},\hat{F}]_S 
&=-[\theta,  e^{-{\cal L}_\Phi}\theta ]_S 
+\frac{1}{2}[e^{-{\cal L}_\Phi}\theta, e^{-{\cal L}_\Phi}\theta ]_S \nn
&=-d_\theta (e^{-{\cal L}_\Phi}\theta ) +\frac{1}{2}e^{-{\cal L}_\Phi} [\theta,\theta]_S \nn
&=-d_\theta (e^{-{\cal L}_\Phi}\theta ),
\ena
where $d_\theta \theta =[\theta,\theta]_S=0$ is used.
Thus, we can regard $\Phi$ as a gauge potential.
The field strength $\hat{F}$ is expanded with respect to $\Phi$ as
\bea
\hat{F} 
&=\sum_{n=1}^\infty \frac{1}{n!} (-{\cal L}_\Phi)^n \theta 
=\frac{e^{-{\cal L}_\Phi} -{\rm id.}}{{\cal L}_\Phi}d_\theta \Phi \nn
&=d_\theta \Phi - \frac{1}{2}[\Phi, d_\theta \Phi ]_S 
+\frac{1}{3!} [\Phi, [\Phi, d_\theta \Phi ]_S ]_S \cdots .
\label{F expansion}
\ena
where ${\cal L}_\Phi \theta =[\Phi,\theta]_S =-d_\theta \Phi$ is used.
The first term $d_\theta \Phi$ is similar to the abelian gauge field strength, but in addition to this term, 
there are infinitely many non-linear corrections.

The gauge transformation of the potential $\Phi$ is defined
 as an insertion of a diffeomorphism generated by a Hamiltonian vector field 
$d_\theta h =X_h \in TM$ for $h \in C^\infty (M)$ as
\bea
e^{-{\cal L}_\Phi}\to  e^{-{\cal L}_{\Phi'}}=e^{-{\cal L}_\Phi}e^{-{\cal L}_{d_\theta h}}.
\label{gauge trf}
\ena
Apparently, $\hat{F}$ in (\ref{F via Phi}) 
is invariant by this transformation, 
since ${\cal L}_{d_\theta h}\theta =-d_\theta^2 h =0$ holds.
Due to the Baker-Campbell-Hausdorff (BCH) formula, the existence of such ${\cal L}_{\Phi'}$ is guaranteed.
The first few terms in $\Phi'$ are obtained by the BCH-formula as
\bea
\Phi' =\Phi +d_\theta h -\frac{1}{2}[\Phi, d_\theta h ]_S 
+ \frac{1}{12}[\Phi, [\Phi, d_\theta h ]_S]_S 
- \frac{1}{12}[d_\theta h, [\Phi, d_\theta h ]_S]_S+\cdots,
\ena
which is again a non-linear extension of the abelian gauge transformation.
It is also possible to exchange the order in (\ref{gauge trf}) as
\bea
e^{-{\cal L}_\Phi}e^{-{\cal L}_{d_\theta h}}
=e^{-{\cal L}_{d_{\theta'} h'}}e^{-{\cal L}_\Phi},
\label{order}
\ena
where $h'= e^{-{\cal L}_{\Phi}} h $.
To show this, we first use a variant of the BCH formula to obtain 
$e^{-{\cal L}_\Phi}e^{-{\cal L}_{d_\theta h}}
=e^{-{\cal L}_{W}}e^{-{\cal L}_\Phi}$ with $W=e^{-{\cal L}_{\Phi}}d_\theta h $.
Next, use the relation:
\bea
d_{\theta'} h' 
&=[\theta + \hat{F}, h' ]_S \nn
&=[e^{-{\cal L}_\Phi}\theta , e^{-{\cal L}_{\Phi}} h]_S \nn
&=e^{-{\cal L}_\Phi} [\theta, h ]_S \nn
&=e^{-{\cal L}_\Phi} d_\theta h .
\ena
In the latter form of the gauge transformation in (\ref{order}), the field strength $\hat{F}$ 
is invariant, since $d_{\theta'} h' $ is a Hamiltonian vector field with respect to $\theta'$:
\bea
\hat{F} &\to e^{-{\cal L}_{d_{\theta'} h'}}e^{-{\cal L}_\Phi}\theta -\theta \nn
&=e^{-{\cal L}_{d_{\theta'}h'}} \theta' -\theta \nn
&=\theta' -\theta \nn
&=\hat{F}.
\ena
Therefore, the gauge transformation associated with the gauge field $\Phi$ preserves 
both $L_\theta$ and $L_{\theta'}$ and 
is generated by a Hamiltonian diffeomorphism.
Note that the set of Hamiltonian vector fields (and thus diffeomorphisms) forms a Lie algebra 
\bea
[{\cal L}_{d_\theta h_1}, {\cal L}_{d_\theta h_2}]
={\cal L}_{[d_\theta h_1, d_\theta h_2]_S}
={\cal L}_{d_\theta \{ h_1,  h_2\}},
\ena
where $\{\cdot, \cdot\}$ is a Poisson bracket with respect to $\theta$.
This shows that the gauge symmetry is not abelian.

In summary, we obtain a gauge theory where the field strength is 
given by the bivector $\hat{F}$, and identified the gauge transformations.
This justifies our discussion on the fluctuation in \S 3.4.
The field strength can be written by the gauge potential, 
a vector field $\Phi$, of which the gauge transformation law is non-nonlinear. 
Interestingly, the Bianchi identity of this gauge theory is the Maurer-Cartan-type relation, which guarantees that the $\beta$-transformed subbundle $L_{\theta+\hat F}$ becomes again the Dirac structure.
Such kind of gauge theory is new (at least to the authors), 
but it is natural from the viewpoint of the differential geometry associated with the Lie algebroid
$(T^* M)_\theta$.

Here, we have seen 
just necessary conditions in constructing a gauge theory.
It is interesting to consider a action functional for $\hat{F}$, but because a Yang-Mills type action 
needs a induced (generalized) metric on the D-brane, we left this problem in the future publication. 
In this paper, we concentrated on to establish the relation between the two gauge theories corresponding to $\tilde F$ and $\hat F$.

\subsubsection{Poisson cohomology}
Finally, we comment on a related mathematical argument.
In \cite{CW}, infinitesimal deformations of a Poisson structures are studied.
They consider a deformation 
\bea
\theta (\epsilon )=\theta +\epsilon \theta_1 +\epsilon^2 \theta_2 +\cdots,
\ena
as a formal power series in a infinitesimal parameter $\epsilon $,
and the condition for $\theta (\epsilon )$ to be a Poisson structure is obtained order by order in $\epsilon $:
\bea
d_\theta \theta_1 =0, \quad 
d_\theta \theta_2 +\frac{1}{2}[\theta_1,\theta_1]_S=0,\cdots
\label{formal deformaiton}
\ena
They are set of recursive equations and related to the second Poisson cohomology group $H^2_\theta$.
In our case, if we insert the parameter $\epsilon $ by replacing $\Phi \to \epsilon \Phi$, 
the expansion in (\ref{F expansion}) directly gives a power series in $\epsilon$. 
Then the first condition in (\ref{formal deformaiton}) is trivially solved by
$\theta_1=d_\theta \Phi $, which means that the deformation is trivial.
The second condition in (\ref{formal deformaiton}) is also solved
by $\theta_2 =-\frac{1}{2}[\Phi, d_\theta \Phi]_S$.
It means that the trivector $[\theta_1,\theta_1]_S$ is $d_\theta$-exact and thus trivial in $H_\theta^3$,
which captures the obstructions to continuing infinitesimal deformations.
The same considerations to higher order equations are needed in this approach.
Instead, we know that our $\theta'= e^{-{\cal L}_{\epsilon \Phi}}\theta$ is a solution to all orders, 
because $\theta'$ is a Poisson structure automatically.
That is, we get an expression for finite trivial deformation $[\theta']=[\theta] \in H_\theta^2$.

To summarize this section, by identifying the 
broken symmetries as fluctuations in two ways, 
we found the two kinds of gauge potentials $a$ and $\Phi$ and associated gauge symmetries.
If they are the two different descriptions of the same Dirac structure of a corresponding D-brane, 
these two gauge fluctuations should also be equivalent. 
So our next task is to find a relation between the 
two gauge fields $a$ and $\Phi$.
To this end, the diffeomorphism appearing in the Moser's lemma plays 
an important role
and thus in the next section we reformulate 
the Moser's lemma in the framework 
of generalized geometry.

\section{Moser's Lemma in Generalized Geometry}

Here we show that the Moser's lemma \cite{Moser:1965} and its Poisson version \cite{Jurco:2001my}
are understood quite naturally within the generalized geometry framework.
As a result, we obtain another expression for the bivector field strength $\hat{F}$.

\subsection{Moser's lemma}

We start with a quick review of Moser's lemma.
Suppose we have a symplectic form $\omega\in \wedge^2 T^*M$, 
$i.e.$ a non-degenerate closed 2-form on $M$, and consider a 
deformation $\omega'=\omega + \tilde{F}$
by adding an exact $2$-form $\tilde{F}=da$.
Moser's lemma states that there exists a diffeomorphism $\rho_a :M\to M$
 which realize the deformation as $\omega'=\rho_a^* \omega$.

Such a diffeomorphism $\rho_a$ is constructed in the following way.
Let us define one-parameter family of symplectic forms by
\begin{align}
	\omega_t =\omega + t da=\omega + t \tilde{F},
	\label{symplectic deformation}
\end{align}
where $t$ is a time parameter $t\in [ 0,1]$, 
and the boundary conditions are $\omega_0=\omega$ and $\omega_1=\omega'$. 
If we define a vector field $X_a(t)$ such that 
\begin{align}
	\omega_t (X_a(t)) = i_{X_a(t)} \omega_t = a,
\end{align}
then the Lie derivative generated by this vector field satisfies the differential equation
\begin{align} 
	\mathcal{L}_{X_a(t)} \omega_t &= d i_{X_a(t)} \omega_t 
	= da = \tilde{F} = \dot{\omega}_t.
\end{align}
Integrating this equation, we obtain the flow 
generated by $X_a(t)$ which relates $\omega$ and $\omega+ \tilde{F}$.

In \cite{Jurco:2001my}, its Poisson version is considered.
They considered a one-parameter deformation of Poisson structure $\theta_t$, 
which is defined by a flow equation
\bea
\dot{\theta}_t &=\cL_{\theta_t (a)} \theta_t 
=d_{\theta_t} \theta_t (a)
=-\theta_t \tilde{F} \theta_t,
\label{flow equation}
\ena
where $\tilde{F}=da$ and the last expression is understood as a matrix multiplication.
Its solution (with initial condition $\theta_0=\theta$) is given by
\bea
\theta_t =\theta \frac{1}{1+t \tilde{F}\theta}.
\label{theta_t}
\ena 
Since the equation (\ref{flow equation}) defines a flow generated by a vector field $\theta_t (a)$, by integrating, we obtain   
a diffeomorphism $\rho_a$ such that $\rho^*_{a} \theta_1= \theta_0$.
It is a Poisson map (symplectomorphism):
\bea
\rho^*_a \{f, g\}_{\theta_1}
=\rho^*_a (\theta_1 (df,dg))
=\theta_0 (d(\rho^*_a f), d(\rho^*_a g))
=\{\rho^*_a f, \rho^*_a g\}_{\theta_0}.
\ena
Note that $X_a(t) =\theta_t(a)$ in the symplectic 
case where $\theta=\omega^{-1}$.

The Moser's lemma relates a change of a symplectic (Poisson) structure 
to a one-parameter diffeomorphisms.
We will see below that it originates from more general relations 
between $B$($\beta$)-transformations and diffeomorphisms.

\subsection{Moser's diffeomorphism}
First, we show that an arbitrary diffeomorphism acting on a Dirac structure $L_\omega=e^\omega (TM)$ 
equals to a $B$-field gauge transformation up to $L_\omega$-diffeomorphism (symmetry).

\begin{figure}[htbp]
\begin{center}
\includegraphics[width=0.4\textwidth]{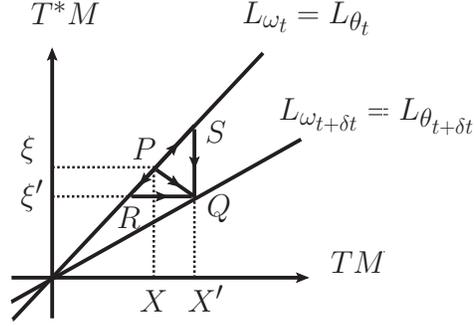}
\caption{Two decompositions of an infinitesimal diffeomorphism acting on $L_{\theta_t}=L_{\omega_t}$}
\label{fig:decomposition}
\end{center}
\end{figure}

To this end, we decompose 
an infinitesimal diffeomorphism transformation 
of a section $X+\omega (X) \in L_\omega $ generated by a vector field $\epsilon \in TM $ as
\bea
&X+\omega (X) \nn
&\xrightarrow{~{\cal L}_\epsilon ~} X+\omega (X) +{\cal L}_\epsilon (X+\omega (X))
=X'+\omega (X') + ({\cal L}_\epsilon \omega ) (X') ,
\ena
up to ${\cal O}(\epsilon^2)$, where $X'=X+{\cal L}_\epsilon X$.
The term $X'+\omega (X') \in L_\omega$ is obtained by a $L_\omega$-diffeo. 
generated by $\epsilon +\omega(\epsilon ) \in L_\omega $.
The term $({\cal L}_\epsilon \omega ) (X')$ is a result of a $B$-gauge transformation.
By noting ${\cal L}_\epsilon \omega =di_\epsilon \omega=d (\omega(\epsilon ))$, 
it is also written by a generalized Lie derivative $-{\cal L}_{\omega(\epsilon )}$ generated by 
a $1$-form $\omega(\epsilon ) \in T^*M$.
Thus, a diffeomorphism can be seen as two-step transformations, 
\bea\
&X+\omega (X) \nn
&\xrightarrow{~{\cal L}_{\epsilon +\omega(\epsilon )}~} X'+\omega (X') \in L_\omega \nn
&\xrightarrow{~-{\cal L}_{\omega(\epsilon )}~} X'+\omega (X')+ ({\cal L}_\epsilon \omega ) (X') ,
\label{diff=B+sym}
\ena
corresponding to the decomposition 
$\epsilon =(\epsilon +\omega(\epsilon ))-\omega (\epsilon )$
or equivalently 
${\cal L}_\epsilon ={\cal L}_{\epsilon +\omega(\epsilon )}-{\cal L}_{\omega (\epsilon )}$.
A graphical explanation is given in Fig.\ref{fig:decomposition} where
 the diffeomorphism represented by the arrow $P \to Q$
is decomposed into the two arrows $P \to S$ and $S \to Q$. 
The arrow $P \to S$ corresponds to a $L_\omega$-diffeomorphism 
preserving the Dirac structure $L_\theta =L_\omega$.
On the other hand, the arrow $S\to Q$ is a $B$-gauge transformation.

Only the $B$-gauge transformation contributes to the net transformation of 
$L_\omega$ as a set of sections.
As a result, the diffeomorphism can be seen as a map 
$L_\omega \to L_{\omega + {\cal L}_\epsilon \omega}$ 
of Dirac structures.

Let us apply this decomposition to the Moser's situation.
In this case, the input is a family of symplectic structures (\ref{symplectic deformation}).
They define a family of Dirac structures $L_{\omega_t}=e^{\omega_t} (TM)$.
The time evolution from $L_{\omega_t}$ to $L_{\omega_{t+\delta t}}$ is 
governed by the $B$-field gauge transformation generated by the difference 
$\omega_{t+\delta t}-\omega_t=\delta t da$.
Then, corresponding to this $B$-field gauge transformation,
a vector field $\epsilon_t$ at time $t$ will be determined through 
the decomposition (\ref{diff=B+sym}).
(In Fig.\ref{fig:decomposition}, 
$P$ and $S$ locate in $L_{\omega_t}$ and $Q$ lies in $L_{\omega_{t+\delta t}}$.
Then, what we have performed here is that for a given $B$-transformation of $S\to Q$, we determine the arrow $P\to Q$.)
Therefore, ${\cal L}_{\epsilon} \omega$  in (\ref{diff=B+sym}) should be 
${\cal L}_{\epsilon_t} \omega_t =d(\omega_t (\epsilon_t))=da$, and 
we obtain the Moser's diffeomorphism $\epsilon_t =\theta_t (a)$ (up to exact $1$-form).
In other words, 
Moser's diffeomorphism ${\cal L}_{\theta_t (a)}$ is decomposed into
${\cal L}_{\theta_t (a)}={\cal L}_{\theta_t (a)+a} -{\cal L}_a$,
which is effectively a $B$-field gauge transformation $-{\cal L}_a$ from the Dirac structure viewpoint.
Here $\theta_t$ is considered as the inverse a map of $\omega_t :TM \to T^*M$, for each $t$. 
In component matrix, it is nothing but (\ref{theta_t}):
\bea
\theta_t =\frac{1}{\omega_t }=\frac{1}{\omega +t\tilde{F}}
=\theta \frac{1}{1+t\tilde{F}\theta}.
\ena

\subsection{Evolution equation on $\theta_t$}
Next, we show that the same diffeomorphism ${\cal L}_\epsilon $ 
can also be decomposed into a $\beta$-transformation and a $L_\theta$-diffeomorphism.
For a section $\xi+\theta (\xi) \in L_\theta =L_\omega$ in another parametrization, 
${\cal L}_\epsilon $  acts as
\bea
&\xi+\theta (\xi) \nn
&\xrightarrow{~{\cal L}_\epsilon ~} \xi+\theta (\xi) +{\cal L}_\epsilon (\xi+\theta (\xi))
=\xi'+\theta (\xi') + ({\cal L}_\epsilon \theta ) (\xi') ,
\ena
up to ${\cal O}(\epsilon^2)$, where $\xi'=\xi+{\cal L}_\epsilon \xi$.
Similar to the previous argument, the term $\xi'+\theta (\xi')$ is obtained by 
a $L_\theta$-diffeomorphism ${\cal L}_{\epsilon +\omega (\epsilon )}$ while the term 
$({\cal L}_\epsilon \theta ) (\xi')$ is a result of a $\beta$-transformation $e^{\beta}$ with $\beta={\cal L}_\epsilon \theta$.
Thus, the transformation by $\epsilon$ is decomposed as
\bea\
&\xi+\theta (\xi) \nn
&\xrightarrow{~{\cal L}_{\epsilon +\omega (\epsilon )}~} \xi'+\theta (\xi')  \in L_\omega \\
&\xrightarrow{~e^{{\cal L}_\epsilon \theta}~} \xi'+\theta (\xi') + ({\cal L}_\epsilon \theta ) (\xi') .
\ena
It is also represented graphically in Fig.\ref{fig:decomposition}, where the diffeomorphism
$P\to Q$ is decomposed into $P \to R$ and $R \to Q $.

Again, the diffeomorphism can effectively be seen as a $\beta$-transformation 
$L_\theta \to L_{\theta +{\cal L}_\epsilon \theta}$ from the Dirac structure viewpoint, 
as a map from a Dirac structure to the other.
This fact was already used in \S 5, to find the gauge potential $\Phi$ corresponding to 
the bivector field strength $\hat{F}$.

Let us now apply the above decomposition to the Moser's situation, again.
For the Moser's diffeomorphism $\epsilon_t =\theta_t (a)$, the parameter $\beta$ is written as
\bea
\beta ={\cal L}_{\theta_t (a)} \theta_t = (\theta_t \wedge \theta_t) (da) 
=-\theta_t \tilde{F}\theta_t,
\ena
where $\theta_t \tilde{F}\theta_t$ denotes 
a bivector, the component of which is given by the matrix product 
$\theta_t^{\mu\nu} \tilde{F}_{\nu\lambda}\theta_t^{\lambda\rho}$.
The time derivative of the equation (\ref{theta_t}) gives 
\bea
\dot{\theta_t} 
=\frac{d}{dt}{\omega_t}^{-1} =-\omega_t^{-1}\dot{\omega_t}\omega_t^{-1}
=-\theta_t \tilde{F} \theta_t.
\ena
Thus, we obtain the flow equation 
\be
\dot{\theta_t}=\beta ={\cal L}_{\theta_t (a)} \theta_t~
\ee
which is the desired equation (\ref{flow equation}).
In particular, $\dot{\theta}_t$ is understood as the parameter for the infinitesimal $\beta$-transformation
at time $t$, like in the previous section, the parameter for the $B$-transformation was $\dot\omega_t=\tilde{F}$.
In Fig.\ref{fig:decomposition}, it gives a relation between an arrow $S \to Q$ and an arrow $R \to Q$.

In summary, the diffeomorphism defined in the Moser's lemma can be
identified either $B$-transformation or $\beta$ transformation up to the generalized 
diffeomorphism preserving the Dirac structure. The differential equation shows that the 
flow is generated by these infinitesimal $B$- or $\beta$-transformations, respectively.

\subsection{$\hat{F}$ from $B$-field gauge transformation}
\begin{figure}[htbp]
\begin{center}
\includegraphics[width=0.4\textwidth]{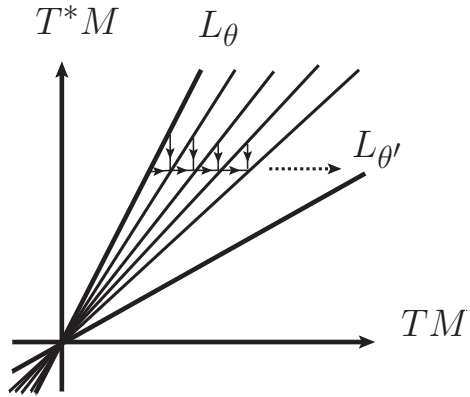}
\caption{A chain of Moser's diffeomorphisms}
\label{fig:GizaGiza}
\end{center}
\end{figure}

We see that the Moser's lemma is used to relate a $B$-field gauge transformation 
and a $\beta$-transformation acting on $L_\theta$.
By integrating (\ref{flow equation}) iteratively, we obtain the formal solution
\bea
\theta'
&=T e^{\int_0^1 dt \cL_{\theta_t (a)}} \theta \nn
&=\sum_{n=1}^\infty \frac{1}{n!}
\int_0^1 \!\!\! dt_1 \!\! \int_0^{t_1} \!\!\! dt_2 \cdots \!\! \int_0^{t_{n-1}} \!\!\!\!\! dt_n 
\cL_{\theta_{t_1} (a)}\cL_{\theta_{t_2} (a)}\cdots \cL_{\theta_{t_n} (a)} \theta.
\ena
where $T$ denotes the symbol of the time ordered product and defined in the second line.
It says that $\theta'$ is obtained as a chain of infinitesimal diffeomorphisms, 
graphically seen in Fig.\ref{fig:GizaGiza}.
Each small triangle in Fig.\ref{fig:decomposition}
corresponds to a decomposition of the infinitesimal diffeomorphism in Fig.\ref{fig:decomposition}.
Since this results in a change of Poisson structures 
$\theta \to \theta'$, the difference $\theta' -\theta$ must give
a bivector $\hat{F}$, 
\bea
\hat{F}=Te^{\int_0^1 {\cal L}_{\theta_t (a)}dt }\theta -\theta.
\ena
In this way,
the original $B$-gauge transformation with the parameter $\tilde{F}=da$
is converted into a $\beta$-transformation $\hat{F}$ through the Moser's diffeomorphism.

\section{Relation between Two Gauge Fields} 

In the previous sections, the two different expressions of the bivector field strength have been obtained:
\bea
\hat{F}=e^{-{\cal L}_\Phi}\theta -\theta, \quad {\rm and}\quad
\hat{F}=Te^{\int_0^1 {\cal L}_{\theta_t (a)}dt }\theta -\theta.
\label{two F hat}
\ena
The purpose of this section is to find a relation between the two gauge potentials $\Phi$ and $a$
from the equivalence of these expressions.
To this end, we need to know the relation between the ordinary and the time-ordered exponentials.
This problem is solved in more wider context, known as the Magnus expansion \cite{Magnus:1954zz}.

\subsection{Magnus expansion}
We summarize necessary information. 
See \cite{Magnus:1954zz} and \cite{Magnus2} for more detail and for applications to other physics.

Consider a differential equation 
\bea
\frac{d}{dt}Y(t) =A(t)Y(t),
\label{linear diff eq}
\ena
with initial condition $Y(0)=Y_0$, 
where $A(t)$ is a matrix (or more generally an operator) valued function of time $t$, and 
$Y(t)$ is a vector valued function to be solved.
Its formal solution is given by the time-ordered exponential as
\bea
Y(t)=Te^{\int_0^t A(s)ds}Y_0.
\label{Texp}
\ena 
Magnus found another representation of the solution of the form
\bea
Y(t)=e^{\Omega(t)}Y_0,
\label{exp}
\ena
that is, a true matrix exponential.
Here $\Omega (t)$ satisfies 
\bea
\dot{\Omega}=\frac{{\rm ad}_\Omega}{e^{{\rm ad}_\Omega} -I}(A)
=\sum_{n=0}^\infty \frac{B_n}{n!}{\rm ad}^n_\Omega (A),
\label{dot omega}
\ena
where the dot denotes the time derivative, 
${\rm ad}_\Omega =[\Omega(t), \cdot]$ and $B_n$ are the Bernoulli numbers.
By integrating it, $\Omega (t)$ is obtained in terms of $A(t)$.

To show that $\Omega(t)$ gives the solution, 
we use an identity of the derivative of matrix exponentials
\bea
\frac{de^{\Omega}}{dt}
=\frac{e^{{\rm ad}_\Omega} -I}{{\rm ad}_\Omega}(\dot{\Omega}) e^{\Omega}
\left(=\int_0^1 e^{s\Omega (t)} \dot{\Omega} e^{(1-s)\Omega (t)}ds \right).
\ena
Then, the l.h.s. of the equation (\ref{linear diff eq}) is written as
\be
\frac{d}{dt}Y(t)
=\frac{de^{\Omega}}{dt}Y_0
=\frac{e^{{\rm ad}_\Omega} -I}{{\rm ad}_\Omega}(\dot{\Omega}) e^{\Omega} Y_0
=\frac{e^{{\rm ad}_\Omega} -I}{{\rm ad}_\Omega}(\dot{\Omega}) Y(t) 
\ee
Using the equation (\ref{dot omega}), we obtain
\be
\frac{d}{dt}Y(t)=\frac{e^{{\rm ad}_\Omega} -I}{{\rm ad}_\Omega}
\left(\frac{{\rm ad}_\Omega}{e^{{\rm ad}_\Omega} -I}(A)\right) Y(t)
=A(t) Y(t)~.
\ee

The exponent $\Omega (t)$ is obtained as an expansion with 
respect to the order of $A(t)$ as 
\bea
\Omega (t)=\sum_{n=1}^\infty \Omega_n (t),
\ena
and is called the Magnus expansion.
The first few terms are obtained 
by substituting the Bernoulli numbers in (\ref{dot omega}) as follows
\bea
\dot{\Omega} (t)=A(t)-\frac{1}{2}[\Omega (t), A(t)] +\frac{1}{12}[\Omega (t),[\Omega (t),A(t)]]+\cdots.
\ena
By comparing the same order in both sides, it leads to the relations
\bea
\dot{\Omega}_1 =A, \quad 
\dot{\Omega}_2 =-\frac{1}{2}[\Omega_1 , A] , \quad 
\dot{\Omega}_3 =-\frac{1}{2}[\Omega_2 , A] +\frac{1}{12}[\Omega_1,[\Omega_1,A]],
\ena
which can be integrated as (we use a notation $A(t_1)=A_1$ ) 
\bea
&\Omega_1 (t)=\int_0^t \!\!\! dt_1 A_1,\nn
&\Omega_2 (t)=\frac{1}{2}\int_0^t \!\!\! dt_1 \!\! \int_0^{t_1} \!\!\! dt_2 [A_1,A_2], \nn
&\Omega_3 (t)=\frac{1}{6}\int_0^t \!\!\! dt_1 \!\! \int_0^{t_1} \!\!\! dt_2 \!\! \int_0^{t_2} \!\!\! dt_3 
\left([A_1,[A_2,A_3]]+[[A_1,A_2],A_3]\right).
\label{omegas}
\ena
See Appendix for the proof, and for a direct check of the equivalence of (\ref{Texp}) and (\ref{exp}) at
a first non-trivial order.

\subsection{Relation using Magnus expansion}
In our context, the Magnus expansion gives an expression of an operator ${\cal L}_\Phi$ 
in terms of a given time-dependent operator ${\cal L}_{\theta_t (a)}$
defined by the equation 
\bea
e^{-{\cal L}_\Phi}=Te^{\int_0^1 {\cal L}_{\theta_t (a)}dt }~.
\label{SW via Magnus}
\ena
As we saw above, the relation is given by the
iterated integrals of multiple commutators
of operators ${\cal L}_{\theta_t (a)}$ as in (\ref{omegas}).
It leads directly to an expression of the vector field $\Phi$ in terms of $\theta_t (a)$, since 
the commutator of Lie derivatives corresponds to the Lie bracket (Schouten bracket) of 
vector fields thanks to $[{\cal L}_X, {\cal L}_Y]={\cal L}_{[X,Y]}$.
By expanding the gauge field $\Phi$ as
\bea
\Phi =\sum_{n=1}^\infty \Phi_n, 
\ena
with respect to the order of $a$, 
each $\Phi_n$ is determined in principle at any order.
First few terms are read off from (\ref{omegas}) as
\bea
&\Phi_1 =-\int_0^1 \!\!\! dt \theta_t (a),\nn
&\Phi_2 =\frac{1}{2}\int_0^1 \!\!\! dt_1 \!\! \int_0^{t_1} \!\!\! dt_2 
[\theta_{t_1} (a),\theta_{t_2} (a)], \nn
&\Phi_3 =-\frac{1}{6}\int_0^1 \!\!\! dt_1 \!\! \int_0^{t_1} \!\!\! dt_2 \!\! \int_0^{t_2} \!\!\! dt_3 
\left([\theta_{t_1} (a),[\theta_{t_2}(a),\theta_{t_3}(a)]]
+[[\theta_{t_1}(a),\theta_{t_2}(a)],\theta_{t_3}(a)]\right).
\label{Phi expansion}
\ena
Thus, the gauge potential $\Phi$ is highly non-linear in $a$.

Under the relation (\ref{SW via Magnus}) between the 
two gauge potentials,
it can be shown that the two gauge theories are gauge equivalent.
That is, given a $U(1)$-gauge transformation $a\to a+d\lambda $ with a gauge parameter (function) $\lambda$, 
there is a gauge parameter $h$ in the other theory 
which defines a gauge transformation (\ref{gauge trf}). 
Such a gauge parameter $h$ can be defined by
\bea
e^{-{\cal L}_\Phi}e^{-{\cal L}_{d_\theta h}}
=Te^{\int_0^1 {\cal L}_{\theta_t (a+d\lambda )}dt }~.
\label{gauge equiv via Magnus}
\ena
As is shown in the proof below,
$h$ is also explicitly determined by the Magnus expansion 
\bea
h=\sum_{n=1}^\infty h_n,
\ena
with respect to the order of $\lambda$.
The first few terms are
\bea
&h_1 =-\int_0^1 \!\!\! dt \tilde{\lambda}(t),\nn
&h_2 =\frac{1}{2}\int_0^1 \!\!\! dt_1 \!\! \int_0^{t_1} \!\!\! dt_2 
[\tilde{\lambda}(t_1),d_\theta\tilde{\lambda}(t_2)], \nn
&h_3 =-\frac{1}{6}\int_0^1 \!\!\! dt_1 \!\! \int_0^{t_1} \!\!\! dt_2 \!\! \int_0^{t_2} \!\!\! dt_3 
\left([\tilde{\lambda}(t_1),[d_\theta\tilde{\lambda}(t_2),d_\theta\tilde{\lambda}(t_3)]]
+[[d_\theta\tilde{\lambda}(t_1),d_\theta\tilde{\lambda}(t_2)],\tilde{\lambda}(t_3)]\right),
\label{first few terms of h}
\ena
where 
\bea
\tilde{\lambda} (t) =e^{{\cal L}_{\Phi (t)}}\lambda,\label{Magnus Gauge}
\ena
with $\Phi (t)$ defined by the Magnus expansion of
$Te^{\int_0^t {\cal L}_{\theta_s (a)}ds }=e^{-{\cal L}_{\Phi(t)}}$.

Before we give the proof, a few remarks are in order.
First, the existence of such $h$ is natural since
$U(1)$-gauge transformation preserves $\tilde{F}$ as well as (\ref{gauge trf}) preserves $\hat{F}$.
However, it is not necessarily trivial that $h$ non-linearly 
depends not only on $\lambda$ but also on the $U(1)$-gauge potential $a$ through $\Phi$.
Note also that the equivalence of (\ref{SW via Magnus})
 is stronger condition than the equivalence of the two expression 
in (\ref{two F hat}).
This means that there are ambiguities in relating $\Phi$ and $a$ 
coming form $\theta$-preserving diffeomorphisms, in addition to the gauge ambiguities.
It is interesting to note that the similarity of these properties with the Seiberg-Witten map, 
i.e., a non-linear relation between the two gauge fields, a gauge equivalence \cite{Seiberg:1999vs},
and its ambiguities \cite{Asakawa:1999cu}.

{\it Proof:} We use an identity valid for any operator-valued function $A(t)$ and $B(t)$ 
of time $t$ :
\be
Te^{\int_0^1 (A(t)+B(t) )dt }=Te^{\int_0^1 A(t)dt }Te^{\int_0^1 B^{(A)} (t) dt }~, 
\label{T-BCH formula}
\ee
where
\be
B^{(A)} (t) =Te^{-\int_0^t A(s)ds }B(t) Te^{\int_0^t A(s)ds }~.
\ee
(the proof of this identity is given in Appendix.)
We apply this identity to the r.h.s. of (\ref{gauge equiv via Magnus})
by setting $A(t)={\cal L}_{\theta_t (a)}$ and $B(t)={\cal L}_{\theta_t (d\lambda )}$.
Then, we have
\bea
B^{(A)} (t) 
&=Te^{-\int_0^t {\cal L}_{\theta_s (a)} ds }
{\cal L}_{\theta_t (d\lambda )} Te^{\int_0^t {\cal L}_{\theta_s (a)}ds }\nn
&=e^{{\cal L}_{\Phi (t)}}{\cal L}_{\theta_t (d\lambda )}e^{-{\cal L}_{\Phi (t)}}\nn
&={\cal L}_{e^{{\cal L}_{\Phi (t)}} \theta_t (d\lambda )}
\ena
Here we used the same $\Phi(t)$ appeared in (\ref{Magnus Gauge}).
Note that $\theta_t = e^{-{\cal L}_{\Phi (t)}}\theta $ is 
corresponding to the Moser's diffeomorphism integrated up to time $t$.
By using the relation $\theta_t (d\lambda )=d_{\theta_t} \lambda =[\theta_t, \lambda]_S$, 
the vector field inside the Lie derivative in the last line becomes
\bea
e^{{\cal L}_{\Phi (t)}} \theta_t (d\lambda )
&=e^{{\cal L}_{\Phi (t)}} [\theta_t, \lambda ]_S \nn
&= [e^{{\cal L}_{\Phi (t)}} \theta_t, e^{{\cal L}_{\Phi (t)}}\lambda ]_S \nn
&= [\theta, e^{{\cal L}_{\Phi (t)}}\lambda ]_S \nn
&=d_\theta (e^{{\cal L}_{\Phi (t)}}\lambda )
\ena
Thus, by defining $\tilde{\lambda} (t)$ as (\ref{Magnus Gauge}), 
the right hand side of (\ref{gauge equiv via Magnus}) is written as
\bea
Te^{\int_0^1 {\cal L}_{\theta_t (a+d\lambda )}dt }
&=Te^{\int_0^1 {\cal L}_{\theta_t (a)}dt } Te^{\int_0^1 {\cal L}_{d_\theta \tilde{\lambda}(t)}dt }.
\ena
Of course, the first term is written as $e^{-{\cal L}_{\Phi }} $ by the Magnus expansion.
By applying the Magnus expansion to the second term, we can also find a vector field $Y$ 
such that 
\bea
Te^{\int_0^1 {\cal L}_{d_\theta \tilde{\lambda}(t)}dt }
=e^{-{\cal L}_{Y}}~.
\ena
First few terms in the expansion $Y=\sum_{n=1}^\infty  Y_n $ w.r.t. the order in $ \tilde{\lambda}(t)$
are read off from (\ref{omegas}) as
\bea
&Y_1 =-\int_0^1 \!\!\! dt d_\theta \tilde{\lambda}(t),\nn
&Y_2 =\frac{1}{2}\int_0^1 \!\!\! dt_1 \!\! \int_0^{t_1} \!\!\! dt_2 
[d_\theta\tilde{\lambda}(t_1),d_\theta\tilde{\lambda}(t_2)], \nn
&Y_3 =-\frac{1}{6}\int_0^1 \!\!\! dt_1 \!\! \int_0^{t_1} \!\!\! dt_2 \!\! \int_0^{t_2} \!\!\! dt_3 
\left([d_\theta\tilde{\lambda}(t_1),[d_\theta\tilde{\lambda}(t_2),d_\theta\tilde{\lambda}(t_3)]]
+[[d_\theta\tilde{\lambda}(t_1),d_\theta\tilde{\lambda}(t_2)],d_\theta\tilde{\lambda}(t_3)]\right).
\ena
As easily seen, all the entries inside the Lie brackets in $Y_n$ $(n \ge 2)$ are $d_\theta$-exact.
Thus, by using the derivation property $d_\theta [X,Y]=[d_\theta X,Y]+[X,d_\theta Y]$ 
and $d_\theta^2=0$ repeatedly,
we can always extract one $d_\theta$
from each $Y_n$.
As a result, $Y$ is $d_\theta$-exact, that is, 
$Y =d_\theta h$ for some $h$.
By construction, the parameter $h$ is also determined as 
the sum $h=\sum_n h_n$, 
and first few terms are 
already given in (\ref{first few terms of h}).

It is also shown that the relation between $a$ and $\Phi$ obtained in this section is compatible
with the relation between $\tilde{F}$ and $\hat{F}$ discussed in \S 3.4.
To this end, recall that the Moser's diffeomorphism relates two symplectic structures as
$\omega'=\omega +\tilde{F}=Te^{\int_0^1 {\cal L}_{\theta_t (a)} dt} \omega$.
By the Magnus expansion, it can be also written as $\omega'=e^{-{\cal L}_\Phi}\omega $.
Now, by applying $e^{-{\cal L}_\Phi}$ on both sides of $\theta \omega=1$,
we have 
\bea
&(e^{-{\cal L}_\Phi}\theta )(e^{-{\cal L}_\Phi}\omega )=e^{-{\cal L}_\Phi} 1 \nn
&\Rightarrow (\theta +\hat{F} )(\omega +\tilde{F} )=1,
\ena
which is nothing but the relation obtained graphically in \S 3.4.

\subsection{Expression using $1$-form}
As a final remark in this section, we point out that the vector field $\Phi$ 
can also be represented by a $1$-form $\hat{A} \in T^*M$ as
\bea
\Phi=\theta (\hat{A}).
\ena
To this end, we first write $\theta_t$ in (\ref{theta_t}) as $\theta_t =\theta Z_t$ with
\bea
Z_t =\frac{1}{1+t \tilde{F}\theta},
\ena
which is a map $Z_t: T^*M \to T^*M$ or equivalently a $(1,1)$-tensor.
Then, we have for example, 
\bea
&\theta_t (a)=\theta (Z_t (a)), \nn
&[\theta_{t_1} (a), \theta_{t_2} (a)]=\theta ( [Z_{t_1}(a), Z_{t_2}(a)]_K),
\ena
where $[\cdot, \cdot]_K$ is the Koszul bracket with respect to $\theta$.
As a result, we find $\hat{A}$, by replacing all the Lie bracket in the Magnus expansion 
(\ref{Phi expansion}) with the Koszul brackets as
\bea
&\hat{A}_1 =-\int_0^1 \!\!\! dt Z_t (a),\nn
&\hat{A}_2 =\frac{1}{2}\int_0^1 \!\!\! dt_1 \!\! \int_0^{t_1} \!\!\! dt_2 
[Z_{t_1} (a),Z_{t_2} (a)]_K, \nn
&\hat{A}_3 =-\frac{1}{6}\int_0^1 \!\!\! dt_1 \!\! \int_0^{t_1} \!\!\! dt_2 \!\! \int_0^{t_2} \!\!\! dt_3 
\left([Z_{t_1} (a),[Z_{t_2}(a),Z_{t_3}(a)]_K]_K
+[[Z_{t_1}(a),Z_{t_2}(a)]_K,Z_{t_3}(a)]_K\right).
\label{hat A expansion}
\ena
In this expression, the new gauge theory based on $L_\theta$ is also written by $1$-form gauge potential.
This is a more similar situation as the relation between commutative and noncommutative gauge theories.

The field strength $\hat{F}$ is also written by $\hat{A}$. 
By definition,
\bea
\hat{F}&=e^{-{\cal L}_{\theta (\hat{A})}}\theta -\theta \nn
&=\sum_{n=1}^\infty \frac{(-1)^n}{n!} {\cal L}^n_{\theta (\hat{A})} \theta.
\ena
Here the first term is rewritten as
\bea
{\cal L}_{\theta (\hat{A})} \theta =(\theta\wedge \theta )(d\hat{A}) 
=-\theta N \theta
\ena
where $N=d\hat{A}$ is a $2$-form and we used the matrix multiplication of tensors in the last expression.
Similarly, the second term becomes
\bea
{\cal L}^2_{\theta (\hat{A})} \theta 
&={\cal L}_{\theta (\hat{A})} (\theta\wedge \theta )(d\hat{A}) \nn
&=({\cal L}_{\theta (\hat{A})} \theta\wedge \theta )(d\hat{A}) 
+ (\theta\wedge {\cal L}_{\theta (\hat{A})} \theta )(d\hat{A}) \nn
&=((\theta\wedge \theta )(d\hat{A}) \wedge \theta )(d\hat{A}) 
+ (\theta\wedge ((\theta\wedge \theta )(d\hat{A})) )(d\hat{A}) \nn
&=2 \theta N \theta N \theta,
\ena
where ${\cal L}_{\theta (\hat{A})} d\hat{A}=0$ is used.
It is easy to find similar expressions for all $n$, and we have
\bea
\hat{F}
&=\sum_{n=1}^\infty \frac{(-1)^n}{n!} (-1)^n n!  \theta N \theta N \cdots N \theta \nn
&=\theta \left( \sum_{n=0}^\infty (N \theta )^n  \right) N \theta \nn
&=\theta \left(\frac{1}{1-N\theta}N \right)\theta.
\ena
Thus, the bivector field strength $\hat{F}$ is always sandwiched by $\theta$. 
The $2$-form field strength $\frac{1}{1-N\theta}N$
is basically given by the $U(1)$ field strength
$N=d\hat{A}$, but dressed with non-linear corrections.

\section{Conclusion and Discussion}

As we have stressed, there are always two ways to specify the same Dirac structure 
in the generalized geometry. 
Since a Dirac structure gives the geometrical characterization of a D-brane, 
these two possibilities gives two types of geometrical descriptions of the D-brane, 
which are equivalent but with quite different appearances. 
In this paper, we have considered two such descriptions of the Dirac structure, 
the one based on the $B$-transformation and the other based on the $\beta$-transformation. 

The characterization of D-branes by Dirac structures is 
very useful for analyzing the fluctuation of D-branes,
since a D-brane with fluctuation is also described by a Dirac structure. 
Thus, to analyze the fluctuation we can use the whole machinery of the deformation theory of the Dirac structure
in the generalized geometry.
In \cite{Asakawa:2012px}, we have already used a part of this theory 
to reveal the full symmetry of D-brane effective theory, the DBI action.
In the present paper, we have made use of this machinery to analyze the gauge symmetry 
of the new representation of the D-brane fluctuation. 
 
We have shown that when considering the bound state of D-branes by assigning
the Dirac structure $L_\omega =L_\theta$,  
we obtain two gauge theories to describe the fluctuations, 
corresponding to the deformation $L_{\omega+\tilde{F}}= L_{\theta+\hat{F}}$
based on the $B$-transformation or the $\beta$-transformation.
The former turns out to be the standard $U(1)$ gauge theory 
with the 2-form field strength $\tilde F$ and a $1$-form gauge potential $a$. 
On the other hand, the latter is a non-standard gauge theory with a bivector field strength $\hat{F}$ 
with the potential of a vector field $\Phi$, and the gauge symmetry is 
diffeomorphisms generated by Hamiltonian vector fields.

Since the two gauge theories are constructed from two different 
deformations of the Dirac structure, they show quite different properties. 
However, when they describe the same fluctuation of the same D-brane, they should be equivalent. 
We found the map
between the two gauge fields, $a$ and $\Phi$, 
by using Moser's lemma and the Magnus expansion. 
The relation between the field strengths, $\tilde{F}$ and $\hat{F}$,
 can be found easily 
by applying a graphical representation of the fluctuation.
On the other hand, to derive the relation between the gauge potentials and the gauge symmetries we need to 
solve the flow equation explicitly. 
The resulting relations are non-linear and formulated as a infinite series. 
We gave a general formula
and the concrete relation for the lower order of the expansion.

Our results have a lot of similarity 
with the typical properties of noncommutative D-branes, 
such as Seiberg-Witten relation for open and closed string metric, 
the SW map for constant fields, 
although our formulas are valid for more general situations.
Note that the existence of such a map between two different gauge fields and their gauge equivalence
are rather natural physically, since they are the properties for any two theories, 
which shares the same S-matrix.
To proceed, it is 
important to verify the equivalence at the action level.
In this paper, we did not discuss about a particular action functional 
for a gauge potential $\Phi$, such as a DBI-type action for $\hat{F}$.
For instance, an Yang-Mills type action functional, which is quadratic in $\Phi$, is highly non-linear in $a$, 
and infinitely many higher derivative terms appear as in the SW map \cite{Jurco:2001my}.
It is also interesting to study whether the equivalence still holds as quantum field theories.

For simplicity, we assumed nondegenerate Poisson structures for $L_\theta$ in this paper. 
However, we can take more general Poisson structures
and can construct a gauge theory with a gauge potential $\Phi$ and a field strength $\hat{F}$ 
also in that case. 
It would correspond to ordinary $U(1)$ gauge theories on symplectic leaves of various dimensions.
The interpretation as D-brane bound state in that case is interesting to study.

One of our original motivations is the understanding of noncommutative description of a D-brane 
in the framework of generalized geometry.
This work is considered as an intermediate step to this end.
What we need to develop is the understanding of the relation between the noncommutative worldvolume and a leaf of foliation, where the latter is embedded into a commutative spacetime.
In the description of Seiberg and Witten \cite{Seiberg:1999vs},
the noncommutativity results from the string worldsheet theory in the constant $B$-field background.
For an arbitrary Poisson manifold, Cattaneo and Felder \cite{Cattaneo:1999fm} 
shows the deep connection 
of the worldsheet theory with the Kontsevich's theory of deformation quantization \cite{Kontsevich:1997vb}.
In Wess et.al. \cite{Jurco:2001my}, 
noncommutative gauge theories are constructed starting from any Poisson manifold.
There, the noncommutativity is realized simply 
by applying the Kontsevich's formality map to gauge theories on Poisson manifolds.
Moreover, they argued that the SW map is related 
to a quantum version of the Moser's diffeomorphism.
These previous works already clarify the mechanism 
of the noncommutativity, however we would like to 
understand it more geometrically within the framework of the generalized geometry.

Note that the formulation of bivector gauge field strength $\hat{F}$ 
is possible only for the Dirac structure $L_\theta$ with non-zero $\theta$.
We do not have any such description on the cotangent bundle $T^*M$.
On the other hand, 
the Dirac structure $L=T^*M$ corresponds to a single $D$-instanton, where its worldvolume is a point.
It is known that infinitely many $D$-instantons can describe a noncommutative D-brane 
\cite{Ishibashi:1999vi, Okuyama:1999ig}.
Therefore, the problem is also related to the understanding of a reliable method to 
treat multiple D-branes in generalized geometry.

Finally, in this paper, we restrict ourselves to the case of vanishing flux
and applied to the D-brane, i.e., to the open string sector.
However the description of Dirac structures developed in this paper would also help to analyze 
a $H$-flux background, or more general non-geometric flux backgrounds in the closed string sector
\cite{Blumenhagen:2011ph, Blumenhagen:2012pc, Blumenhagen:2012nt}. 
Especially, we would like to investigate the role of this new type of gauge theory 
in this context. 
We hope to report on this subject in the near future.

\section*{Acknowledgement}
Authors would like to thank 
the members of the particle theory and cosmology group, 
in particular  S.~Sasa and U.~Carow-Watamura for helpful comments and discussions. 
T.~A. and H.~M. are supported partially by the GCOE program ``Weaving Science Web beyond Particle-Matter Hierarchy'' 
at Tohoku University.

\appendix
\section{Proof of the formula (\ref{key relation for Poisson Dirac})}

Here we prove the identity
\bea
[\theta(\xi),\theta(\eta)]=\theta([\xi,\eta]_K) +\frac{1}{2}[\theta,\theta]_S (\xi,\eta),
\label{to be shown}
\ena
for an arbitrary bivector $\theta \in \wedge^2 TM$.
In particular, if $\theta$ is a Poisson bivector, (\ref{key relation for Poisson Dirac}) holds.

Here we demonstrate it locally in terms of components.
The l.h.s. of (\ref{to be shown}) is written locally by using vector fields 
$\theta^\mu:=\theta^{\mu\nu}\partial_\nu$ as
\bea
[\theta(\xi),\theta(\eta)]
&=[\theta^{\mu\nu}\xi_\mu \partial_\nu ,\theta^{\rho\tau}\eta_\rho \partial_\tau]
=[\xi_\mu \theta^{\mu},\eta_\rho\theta^{\rho}]\nn
&=\xi_\mu \theta^\mu(\eta_\rho)\theta^\rho +\xi_\mu\eta_\rho [\theta^\mu,\theta^\rho]
-\theta^\rho (\xi_\mu)\eta_\rho \theta^\mu,
\ena
where $\theta^\mu(\eta_\rho) =\theta^{\mu\nu}\partial_\nu\eta_\rho$
and $[\theta^\mu,\theta^\rho] = (\theta^{\mu\nu}\partial_\nu \theta^{\rho \tau }
-\theta^{\rho \nu}\partial_\nu \theta^{\mu\tau })\partial_\tau $.
The first term in the r.h.s. of (\ref{to be shown}) is written as
\bea
\theta([\xi,\eta]_K) 
&=\theta^{\rho\tau} \left( 
\theta^{\mu\nu} \xi_\mu \partial_\nu \eta_\rho
+\theta^{\mu\nu} \partial_\mu \xi_\rho \eta_\nu 
+\partial_\rho \theta^{\mu\nu} \xi_\mu \eta_\nu
\right) \partial_\tau \nn
&=\xi_\mu \theta^\mu (\eta_\rho) \theta^\rho
- \theta^\nu (\xi_\rho )\eta_\nu \theta^\rho
+\xi_\mu \eta_\nu \theta^{\rho\tau}\partial_\rho \theta^{\mu\nu} \partial_\tau
\ena
where (\ref{Poisson Lie bracket in components}) is used.
Thus, the difference is obtained as 
\bea
[\theta(\xi),\theta(\eta)]-\theta([\xi,\eta]_A) 
&=\xi_\mu\eta_\rho [\theta^\mu,\theta^\rho] 
-\xi_\mu \eta_\nu \theta^{\rho\tau}\partial_\rho \theta^{\mu\nu} \partial_\tau \nn
&= \xi_\mu\eta_\rho  (\theta^{\mu\nu}\partial_\nu \theta^{\rho \tau }
-\theta^{\rho \nu}\partial_\nu \theta^{\mu\tau }
-\theta^{\nu\tau}\partial_\nu \theta^{\mu\rho})\partial_\tau \nn
&= \xi_\mu\eta_\rho  (\theta^{\mu\nu}\partial_\nu \theta^{\rho \tau }
+\theta^{\rho \nu}\partial_\nu \theta^{\tau \mu}
+\theta^{\tau\nu}\partial_\nu \theta^{\mu\rho})\partial_\tau \nn
&=\frac{1}{2}[\theta,\theta]_S (\xi,\eta),
\ena
which is the desired result (\ref{to be shown}).

\section{On Magnus expansion}

In this section, we prove (\ref{omegas}) in the Magnus expansion, first.
Then, we check the equivalence of (\ref{Texp}) and (\ref{exp}).

\paragraph{The proof of (\ref{omegas})}
The first two equations in (\ref{omegas}) are obvious.
In the third one, we calculate as
\bea
\Omega_3 (t)
&=\int_0^t \!\!\! dt_1 \left( 
-\frac{1}{2}[\Omega_2 (t_1) , A_1] +\frac{1}{12}[\Omega_1 (t_1),[\Omega_1 (t_1),A_1]]
\right) \nn
&=-\frac{1}{4}\int_0^t \!\!\! dt_1 \!\! \int_0^{t_1} \!\!\! dt_2 \!\! \int_0^{t_2} \!\!\! dt_3
[[A_2,A_3],A_1]
+\frac{1}{12}\int_0^t \!\!\! dt_1 \!\! \int_0^{t_1} \!\!\! dt_2 \!\! \int_0^{t_1} \!\!\! dt_3
[A_2, [A_3 , A_1]]\nn
&=-\frac{1}{4}\int_0^t \!\!\! \int_0^t \!\!\! \int_0^t \!\!\! dt_1 dt_2 dt_3 \Theta_{123}
[[A_2,A_3],A_1]
+\frac{1}{12}\int_0^t \!\!\! \int_0^t \!\!\! \int_0^t \!\!\! dt_1 dt_2 dt_3 \Theta_{123}
([A_2, [A_3 , A_1]] + [A_3, [A_2 , A_1]])\nn 
&=\int_0^t \!\!\! \int_0^t \!\!\! \int_0^t \!\!\! dt_1 dt_2 dt_3 \Theta_{123}
\left( \frac{1}{4}[A_1,[A_2,A_3]] +\frac{1}{12}([A_3,[A_2,A_1]]-[A_1, [A_2 , A_3]] +[A_3, [A_2 , A_1]])
\right)\nn
&=\int_0^t \!\!\! \int_0^t \!\!\! \int_0^t \!\!\! dt_1 dt_2 dt_3 \Theta_{123}
\left( \frac{1}{6}[A_1,[A_2,A_3]] +\frac{1}{6}[A_3,[A_2,A_1]]
\right)
\label{3rd order proof}
\ena
Here $\Theta_{123}=\Theta (t_1-t_2)\Theta(t_2-t_3)$ and $\Theta_{12}=\Theta (t_1-t_2)$ is the Heviside 
step function.
Note that $\Theta_{12}\Theta_{23}=\Theta_{12}\Theta_{13}\Theta_{23}$.
In the second term of third line, we convert the ordinary double integral into the time-ordered one using 
the identity 
\bea
\int \!\! dt_1 A_1 \int \!\! dt_2 A_2 \cdots \int \!\! dt_n A_n
=  \int \!\!\! \int \!\!\! \cdots \!\!\! \int \!\! dt_1 dt_2 \cdots dt_n \Theta_{12\cdots n}
\sum_{\sigma \in S_n} A_{\sigma (1)} A_{\sigma (1)}\cdots A_{\sigma (n)},
\label{permutation identity}
\ena
where $S_n$ is the permutation group.
In the fourth line, the Jacobi identity $[A_2, [A_3 , A_1]]=[A_3,[A_2,A_1]]-[A_1, [A_2 , A_3]]$ is used.

\paragraph{Equivalence of (\ref{Texp}) and (\ref{exp})}
With respect to the Magnus expansion, matrix exponential is also expanded as
\bea
e^\Omega =e^{\sum_{n}\Omega_n} 
=1+ \Omega_1 + \left( \Omega_2 + \frac{1}{2}\Omega_1^2\right) +\cdots,
\ena
according to the order of $A(t)$ in the time-ordered exponential.
The equivalence up to the second order is easy to show.
In the third order, we explicitly show the equivalence:
\bea
\Omega_3 +\frac{1}{2} (\Omega_1 \Omega_2 +\Omega_2 \Omega_1)+\frac{1}{6}\Omega_1^3
=\int_0^t \!\!\! \int_0^t \!\!\! \int_0^t \!\!\! dt_1 dt_2 dt_3 \Theta_{123} A_1 A_2 A_3.
\label{3rd order eq}
\ena
First by using (\ref{permutation identity}), 
the term $\Omega_1^3$ in (\ref{3rd order eq}) is rewritten as
\bea
\Omega_1^3 &=
\int_0^t \!\!\! \int_0^t \!\!\! \int_0^t \!\!\! dt_1 dt_2 dt_3 \Theta_{123}
(A_1 \{ A_2, A_3\} +A_2 \{ A_3, A_1\} + A_3 \{ A_1, A_2\}) .
\ena
On the other hand, the integrand of $\Omega_3$ in (\ref{3rd order proof}) 
is
\bea
&[A_1,[A_2,A_3]]+[[A_1,A_2],A_3] \nn
=& 2(A_1 A_2 A_3 +A_3 A_2 A_1) -(A_1 A_3 A_2 +A_2 A_3 A_1 
+A_2 A_1 A_3 +A_3 A_1 A_2).
\ena
Thus, the sum of these two terms in (\ref{3rd order eq}) is simply
\bea
\Omega_3 +\frac{1}{6}\Omega_1^3 &=
\frac{1}{2} \int_0^t \!\!\! \int_0^t \!\!\! \int_0^t \!\!\! dt_1 dt_2 dt_3 \Theta_{123}
(A_1 A_2 A_3 +A_3 A_2 A_1)~ .
\ena

Next, we rewrite the second term in (\ref{3rd order eq}).
Using (\ref{permutation identity}) again, we have
\bea
\Omega_2 \Omega_1 &=
\frac{1}{2} \int_0^t \!\!\! \int_0^t \!\!\! \int_0^t \!\!\! dt_1 dt_2 dt_3 \Theta_{123}
(\Theta_{12}[A_1,A_2]A_3 + \Theta_{23}[A_2,A_3]A_1 \cdots )\nn
&=\frac{1}{2} \int_0^t \!\!\! \int_0^t \!\!\! \int_0^t \!\!\! dt_1 dt_2 dt_3 \Theta_{123}
([A_1,A_2]A_3 +[A_2,A_3]A_1 +[A_1,A_3]A_2)
\ena
where the half of terms in the summation are dropped due to the product of two step functions.
The $\Omega_2 \Omega_1$ can be represented similarly.
Then, we have 
\bea
&\frac{1}{2} (\Omega_1 \Omega_2 +\Omega_2 \Omega_1) \nn
=&\frac{1}{4} \int_0^t \!\!\! \int_0^t \!\!\! \int_0^t \!\!\! dt_1 dt_2 dt_3 \Theta_{123}
(\{[A_1,A_2],A_3\} +\{[A_2,A_3],A_1\} +\{[A_1,A_3],A_2\}) \nn
=&\frac{1}{2} \int_0^t \!\!\! \int_0^t \!\!\! \int_0^t \!\!\! dt_1 dt_2 dt_3 \Theta_{123}
(A_1 A_2 A_3 -A_3 A_2 A_1),
\ena
where
\bea
\{[A_1,A_2],A_3\} +\{[A_2,A_3],A_1\} +\{[A_1,A_3],A_2\}
=A_1 A_2 A_3 -A_3 A_2 A_1.
\ena
(\ref{3rd order eq}) is now obvious.

\section{Time-ordered BCH formula}

Here we show the identity used in (\ref{T-BCH formula})
\bea
Te^{\int_0^1 (A(t)+B(t) )dt }=Te^{\int_0^1 A(t)dt }Te^{\int_0^1 B^{(A)} (t) dt }, \quad
B^{(A)} (t) =Te^{-\int_0^t A(s)ds }B(t) Te^{\int_0^t A(s)ds },
\label{T-BCH formula2}
\ena
which is a time-ordered version of the BCH formula.

For convenience, we denote $T_M (t)$ as
\bea
	T_M (t) = T\text{e}^{\int^t_0 ds M(s)},
\ena
for a time-dependent operator $M(s)$.
This is a formal solution of the differential equation
\bea
	\frac{d}{dt} T_M(t) = M(t) T_M(t),
\label{diff eq for T_M}
\ena
with $T_M (0)=1$.
With this notation, the formula (\ref{T-BCH formula2}) is expressed as
\bea
	T_{A+B}(1) = T_A(1) T_{B^{(A)}}(1)~,
\label{T-BCHsimple}
\ena 
and we prove this below.
First, by using the differential equation (\ref{diff eq for T_M}), we have
\bea
\frac{d}{dt} \left[ T^{-1}_A (t)T_{A+B} (t)\right]
&= \left[ -T^{-1}_A (t) A(t)\right] T_{A+B} (t) +T^{-1}_A (t) \left[ (A(t) +B(t)) T_{A+B} (t) \right]\\
&= T^{-1}_A  (t) B(t) T_{A+B} (t) \\
&=\left[T^{-1}_A  (t) B (t) T_A (t)\right] T^{-1}_A (t) T_{A+B}(t) \\
&=B^{(A)} (t) \left[T^{-1}_A (t) T_{A+B} (t)\right].
\ena
Since this is again the form of (\ref{diff eq for T_M}), we conclude 
\bea
	T^{-1}_A (t) T_{A+B} (t)= T_{B^{(A)}} (t)~,
\ena
and this gives (\ref{T-BCHsimple}) if we put $t=1$.





\end{document}